%% file: CrossoverQuantumToClassical.tex
\documentclass[prl,twocolumn,amsmath,amssymb,floatfix,footinbib,bibnotes,longbibliography]{revtex4-1}
\pdfoutput=1

\usepackage[colorlinks=true,citecolor=blue,linkcolor=magenta]{hyperref}
\usepackage{amsmath}
\usepackage{graphicx}
\usepackage{changepage}
\usepackage{fancyhdr}
\usepackage{amsthm, amssymb}
\usepackage{floatflt}
\usepackage{float}
\usepackage{array}
\usepackage[usenames,dvipsnames]{color}
\usepackage{mathtools}
\usepackage[normalem]{ulem}

\def \mr{\mathrm}

\input{definitions}

\newcommand {\apgt} {\ {\raise-.5ex\hbox{$\buildrel>\over\sim$}}\ }
\newcommand {\aplt} {\ {\raise-.5ex\hbox{$\buildrel<\over\sim$}}\ }

\newcommand {\rem}[1]{}

\def  \w{\omega}
\def  \qinv{Q^{-1}}
\def  \G{\Gamma}
\def  \q0{\frac{\w_k^2}{\G}+z}
\def  \qe{q_{\text{EA}}}
\def  \qd{q_d}
\DeclareMathOperator{\sech}{sech}

\def \titlename {Quantum to classical crossover in many-body chaos and scrambling from relaxation in a glass}
\def \authornames{Surajit Bera$^1$, Venkata Lokesh K. Y$^{2}$ and Sumilan Banerjee$^1$}
\def \affiliations{$^1$Centre for Condensed Matter Theory, Department of Physics, Indian Institute 
	of Science, Bangalore 560012, India\\
	$^2$Centre of High Energy Physics, Indian Institute of Science, Bangalore 560012, India}

\begin{document}
	
	\title{\titlename}
	\author{\authornames}
	\affiliation{\affiliations}
	\email{surajit@iisc.ac.in}
	\email{venkatakumar@iisc.ac.in}
	\email{sumilan@iisc.ac.in}
	\date\today

\begin{abstract}
	Chaotic quantum systems with Lyapunov exponent $\lambda_\mathrm{L}$ obey an upper bound $\lambda_\mathrm{L}\leq 2\pi k_\mathrm{B}T/\hbar$ at temperature $T$, implying a divergence of the bound in the classical limit $\hbar\to 0$. Following this trend, does a quantum system necessarily become `more chaotic' when quantum fluctuations are reduced? Moreover, how do symmetry breaking and associated non-trivial dynamics influence the interplay of quantum mechanics and chaos? We explore these questions by computing $\lambda_\mathrm{L}(\hbar,T)$ in the quantum spherical $p$-spin glass model, where $\hbar$ can be continuously varied. We find that quantum fluctuations, in general, make paramagnetic phase less and the replica symmetry-broken spin glass phase more chaotic. We show that the approach to the classical limit could be non-trivial, with non-monotonic dependence of $\lambda_\mathrm{L}$ on $\hbar$ close to the dynamical glass transition temperature $T_d$.  Our results in the classical limit ($\hbar\to 0$) naturally describe chaos in super-cooled liquid in structural glasses.
	We find a maximum in $\lambda_\mathrm{L}(T)$ substantially above $T_d$, concomitant with the crossover from simple to slow glassy relaxation. We further show that  $\lambda_\mathrm{L}\sim T^\alpha$, with the exponent $\alpha$ varying between 2 and 1 from quantum to classical limit, at low temperatures in the spin glass phase. 

\end{abstract}

\maketitle 
Understanding thermalization and transport rates in many-body systems and how quantum mechanics affects these rates across various phases and phase transitions have important implications for a remarkably wide range of topics. These include information scrambling in black holes \cite{Sekino2008,Maldacena2015} and quantum circuits \cite{Hosur2016}, strange metals and Planckian dissipation \cite{Bruin2013,Hartnoll2014} and complex dynamics in disordered systems \cite{Berthier2011,Abanin2019}. Recently a quantum Lyapunov exponent or scrambling rate $\lambda_\mathrm{L}$ \cite{Larkin1969} has emerged as one of the important diagnostics of thermalization for  several important systems \cite{Maldacena2015,KitaevKITP,Kitaev2018,Aleiner2016,Patel2017} in high-energy and condensed matter physics. Quantum mechanics fundamentally influences this quantity by setting an upper bound $\lambda_\mathrm{L}\leq 2\pi k_\mathrm{B}T/\hbar$~\cite{Maldacena2015} for a system at temperature $T$. 

However, typically the Lyapunov exponent can only be extracted for quantum systems with a suitable semiclassical limit \cite{Maldacena2015,KitaevKITP,Kitaev2018,Aleiner2016,Patel2017}. An important class of models for such systems corresponds to the solvable large-$N$ Sachdev-Ye-Kitaev (SYK) model \cite{Sachdev2015,KitaevKITP,Kitaev2018} and its variants \cite{Gu2016,BanerjeeAltman2016,Davison2016thermo,SJian2017a,Haldar2018}, where $\lambda_\mathrm{L}$ can be calculated exactly in the large-$N$ semiclassical limit. Nevertheless, once this limit is taken, no other quantum parameter like `$\hbar$' can be tuned in the SYK-type models to explore how quantum mechanics actually intervenes in the evolution of chaos between the classical and quantum limits. Also these models typically do not exhibit any symmetry breaking phase transitions and associated non-trivial dynamics. To address these, we study many-body chaos in one of the most studied solvable models of glasses, namely the spherical $p$-spin glass model \cite{Derrida1980,Kirkpatrick1987,Crisanti1992,Nieuwenhuizen1995,Nieuwenhuizen1995a,Cugliandolo1998,Cugliandolo1999,Cugliandolo2000,Cugliandolo2000,Cugliandolo2001,Reichman2019}. We show that the $p$-spin glass model gives us highly tunable analytical access to the interplay between chaos, quantum fluctuations, symmetry breaking and complex dynamics. 

\begin{figure}[h!]
	\centering
	\includegraphics[width=1.0\linewidth]{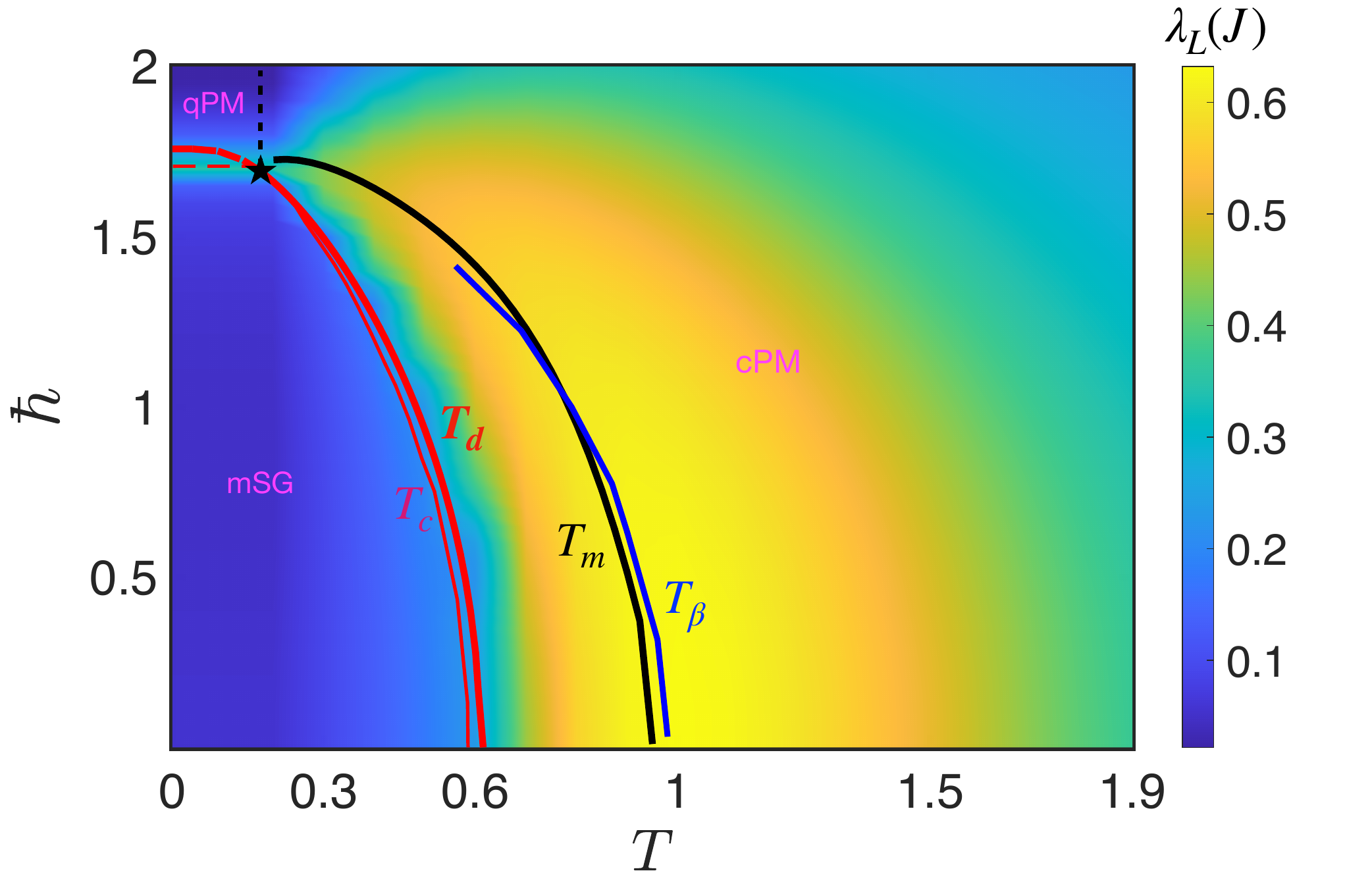}
	\caption{Lyapunov exponent $\lambda_\mathrm{L}(\hbar,T)$ (colormap, in units of $J$) on the thermodynamic phase diagram. The thermodynamic PM-SG phase transition $T_c(\hbar)$ line (thin red solid line) is second order up to a tri-critical point (black star) and then first order (thin dashed red line). The mSG to PM transition is demarcated by the dynamical transition line $T_d(\hbar)>T_c(\hbar)$  (thick solid red line). The locus of the broad maximum of $\lambda_\mathrm{L}(T, \hbar)$ is shown as $T_m(\hbar)$ line (solid black line) and compared with the crossover temperature $T_\beta(\hbar)$ to the two-step glassy relaxation regime (blue line).}
	\label{fig:PhaseDiagram}
\end{figure}

We compute the Lyapunov exponent in the quantum $p$-spin glass model \cite{Nieuwenhuizen1995,Nieuwenhuizen1995a,Cugliandolo1998,Cugliandolo1999,Cugliandolo2000,Cugliandolo2000,Cugliandolo2001} of $N$ spins interacting with random all-to-all $p$-spin interactions. The model shares many common features with other models of quantum spin glass, like transverse-field models \cite{Goldschmidt1990,Dobrosavljevic1990,Nieuwenhuizen1998,Obuchi2007,Nieuwenhuizen1998}. Unlike the latter, the quantum $p$-spin glass model is solvable both in the classical and quantum limits for $N\to\infty$. Moreover, the dynamics of the model in the classical limit $\hbar\to 0$ is of great importance for structural glasses \cite{Berthier2011} and is identical  to the mode coupling theory (MCT) dynamics in super-cooled liquids \cite{GotzeBook,Kob1997,Reichman2005}. As shown in Fig.\ref{fig:PhaseDiagram}, the model has thermodynamic transition, $T_c(\hbar)$, between paramagnetic (PM) and replica-symmetry broken (RSB) spin glass (SG) phase for $p\geq 3$ \cite{Cugliandolo2000,Cugliandolo2001}. There is a dynamical transition at $T_d>T_c$ from slow glassy thermalization to lack of ergodicity below $T_d$ and a relaxation time $\tau_\alpha$, extracted from spin-spin correlation function, diverges for $T\to T_d^+$.

 We obtain $\lambda_\mathrm{L}(\hbar,T)$ from the out-of-time-ordered correlator (OTOC) \cite{Larkin1969,Maldacena2015} for the quantum spin glass by varying $\hbar$ over the entire phase diagram [Fig.\ref{fig:PhaseDiagram}], with the following main results. \\
 1. We show that quantum fluctuations, in general, reduce chaos in the disordered phase (PM) and increase chaos in the ordered (SG) phase. However, we find that $\lambda_\mathrm{L}$, over certain temperature range close to $T_d(\hbar)$ in the PM phase, is a non-monotonic function of $\hbar$. This indicates non-trivial nature of quantum corrections to $\lambda_\mathrm{L}$. \\
2. By taking $\hbar\to 0$ limit for $T>T_d$, we obtain the temperature dependence of the Lyapunov exponent of a super-cooled liquid.\\
3. We show that, unlike $\tau_\alpha$, $\lambda_L^{-1}$ has a broad minimum at $T=T_m>T_d$ [Fig.\ref{fig:PhaseDiagram}], correlated with the crossover to the two-step glassy relaxation \cite{GotzeBook,Kob1997,Reichman2005}. We analytically show that $T_m$ signifies a crossover in chaos, arising due to an interplay of relaxation, the rapid increase of relaxation time in the glassy regime, and the crossover from strong coupling ($\gtrsim T$) to weak coupling ($\lesssim T$). This result is more general than the model considered here and should have implications for complex relaxations in liquids and many other interacting systems.\\
4. For $\lambda_\mathrm{L}$ in the SG phase, we obtain the OTOC in a replica-symmetry broken marginal SG (mSG) phase \cite{Cugliandolo2000,Cugliandolo2001}. 
 We find $\lambda_\mathrm{L}\sim T^\alpha$ at low temperature in the mSG phase, with the exponent $\alpha$ varying between $\sim 2-1$ from quantum to the classical limit.  

Earlier works \cite{Mao2020,Cheng2019} have studied chaos in the PM phase of a quantum rotor glass model \cite{Ye1993} with 2-rotor interaction. The model has the same thermodynamic phase diagram as the $p=2$ spin glass model and the SG phase is replica symmetric \cite{Kosterlitz1976} . Similar SG phase is also realized in a version of SYK model represented in terms of $SO(N)$ spins, where the Lyapunov exponent has been computed via numerical simulation in the classical large spin limit \cite{Scaffidi2019}. 



{\it Model.---}  We study the quantum spherical $p$-spin glass model \cite{Cugliandolo1998,Cugliandolo1999,Cugliandolo2000,Cugliandolo2000,Cugliandolo2001}, described by the Hamiltonian,
\begin{align}
\mathcal{H} &= \sum_i\frac{\pi_i^2}{2M} + \sum_{p,i_1<...<i_p}J_{i_1...i_p}^{(p)}s_{i_1}...s_{i_p}, \label{eq:Hamiltonian}
\end{align}
with random all-to-all interactions among $p=2,3,\dots$ spins on $i=1,\dots,N$ sites; the couplings  $J_{i_1...i_p}^{(p)}$s drawn from Gaussian distribution with variance $J^2_p p!/2N^{p-1}$. The quantum dynamics results from the commutation relation $[s_i, \pi_j ]= \ci \hbar \delta_{ij}$. The model is non-trivial due to the spherical constraint $\sum_i s^2_i = N$. The Hamiltonian describes a particle with mass $M$ moving on the surface of an $N$-dimensional hypersphere. 
We study chaos in the model with $p=3$ ($J_3=J$). For $p=2$, the model is non-interacting \cite{Kosterlitz1976} and non-chaotic, i.e. $\lambda_\mr{L}=0$. 

{\it Large-$N$ saddle points and phase diagram.---} The equilibrium and dynamical phase diagrams of the model [Eq.\eqref{eq:Hamiltonian}] have been analyzed in detail~\cite{Nieuwenhuizen1995,Nieuwenhuizen1995a,Cugliandolo1998,Cugliandolo1999,Cugliandolo2000,Cugliandolo2000,Cugliandolo2001,Thomson2020}. In the $N\to\infty$ limit, the phases are characterized by disorder averaged time-ordered ($\mathcal{T}_\tau$) correlation function, $Q_{ab}(\tau)=(1/N)\sum_i\langle \mathcal{T}_\tau s_{ia}(\tau)s_{ib}(0)\rangle$, obtained from the saddle point equations of the imaginary time ($\tau$) path integral [see Supplementary Material (SM), Sec. \textcolor{red}{S1}],
\begin{subequations} \label{eq:SaddlePoint}
\begin{align}
\qinv_{ab}(\w_k) = \left(\q0\right)\delta_{ab} - \Sigma_{ab}(\w_k) \\
\Sigma_{ab}(\tau) =\sum_p\frac{p \tilde{J}_p^2}{2}[Q_{ab}(\tau)]^{p-1}.
\end{align}
\end{subequations}
The replicas $a=1,..,n$ are introduced to perform the disorder averaging and $\w_k=2k\pi T$ is bosonic Matsubara frequency with $k$ an integer ($k_\mr{B}=1$); $\tilde{J}_p=J_p/J$, and temperature, time and frequency are in units of $J$, $\hbar/J$ and $J/\hbar$, respectively. $Q_{ab}(\omega_k)=\int_0^{\beta}d\tau e^{\ci \omega_k\tau}Q_{ab}(\tau)$ ($\beta=1/T$) 
is matrix in replica space and the spherical constraint, $(1/N)\sum_i s_{ia}^2=Q_{aa}(\tau=0)=1$, is imposed via the Lagrange multiplier $z$. The quantum fluctuations is tuned through the dimensionless parameter $\Gamma=\hbar^2/MJ$ by changing $\hbar$ with fixed $M$ \cite{footnote1}. 

As in the earlier works \cite{Cugliandolo2000,Cugliandolo2001}, we obtain the phase diagram [Fig.\ref{fig:PhaseDiagram}] by numerically solving the saddle-point equations [Eqs.\eqref{eq:SaddlePoint}] (see SM, Sec. \textcolor{red}{S1 1, S1 3}). The replica structure of $Q_{ab}(\tau)$ for $n\to 0$ characterizes PM and SG phases, namely --- (a) in the {\it PM phase},   $Q_{ab}(\tau)=Q(\tau)\delta_{ab}$ is replica symmetric, and (b) for the {\it SG phase}, the order parameter has an exact one-step replica symmetry breaking (1-RSB) structure where $n$ replicas are broken into diagonal blocks with $m$ replicas and  $Q_{ab}(\tau)=(q_d(\tau)-q_\mr{EA})\delta_{ab}+q_\mr{EA}\epsilon_{ab}$; $\epsilon_{ab}=1$ if $a, b$ are in diagonal block else $\epsilon_{ab}=0$. The Edward-Anderson (EA) order parameter $q_\mr{EA}$ is finite in the SG phase and vanishes in the PM phase. 

As shown in Fig.\ref{fig:PhaseDiagram}, the PM to SG phase transition $T_c(\hbar)$ is second order up to a tricritical point and then first order till $T=0$ \cite{Cugliandolo2000,Cugliandolo2001}. 
There are two PM phases, a classical PM (cPM), adibatically connected to PM at $\hbar=0$, and a quantum PM (qPM) phase for $T\lesssim T^{*}$ and above the first-order line. We compute $\lambda_\mr{L}$ in the cPM region since the qPM is strongly gapped \cite{Cugliandolo2001}, and hence very weakly chaotic. For the SG phase, we only consider the so-called marginal spin glass phase \cite{Cugliandolo2001}, where the block size or the break point $m$ is obtained by the \emph{marginal stability criterion} \cite{Cugliandolo2001} (see SM, Sec. \textcolor{red}{S1 2}). The mSG phase is demarcated by the dynamical phase transition line $T_d(\hbar)>T_c(\hbar)$ [Fig.\ref{fig:PhaseDiagram}]. 


For computing the Lyapunov exponent $\lambda_\mathrm{L}$, we also need dynamical correlation and response functions in real time (frequency) $t$ ($\omega$). These are obtained using the spectral function $\rho(\omega)=-\mathrm{Im}Q_{aa}^R(\omega)/\pi$, where the retarded propagator, $Q^R_{ab}(\omega)=Q_{ab}(i\omega_k\to\omega+\ci 0^+)=Q^R(\omega)\delta_{ab}$ (SM, Sec. \textcolor{red}{S1 1}).


\begin{figure}[h!]
	\centering
	\includegraphics[width=1.0\linewidth]{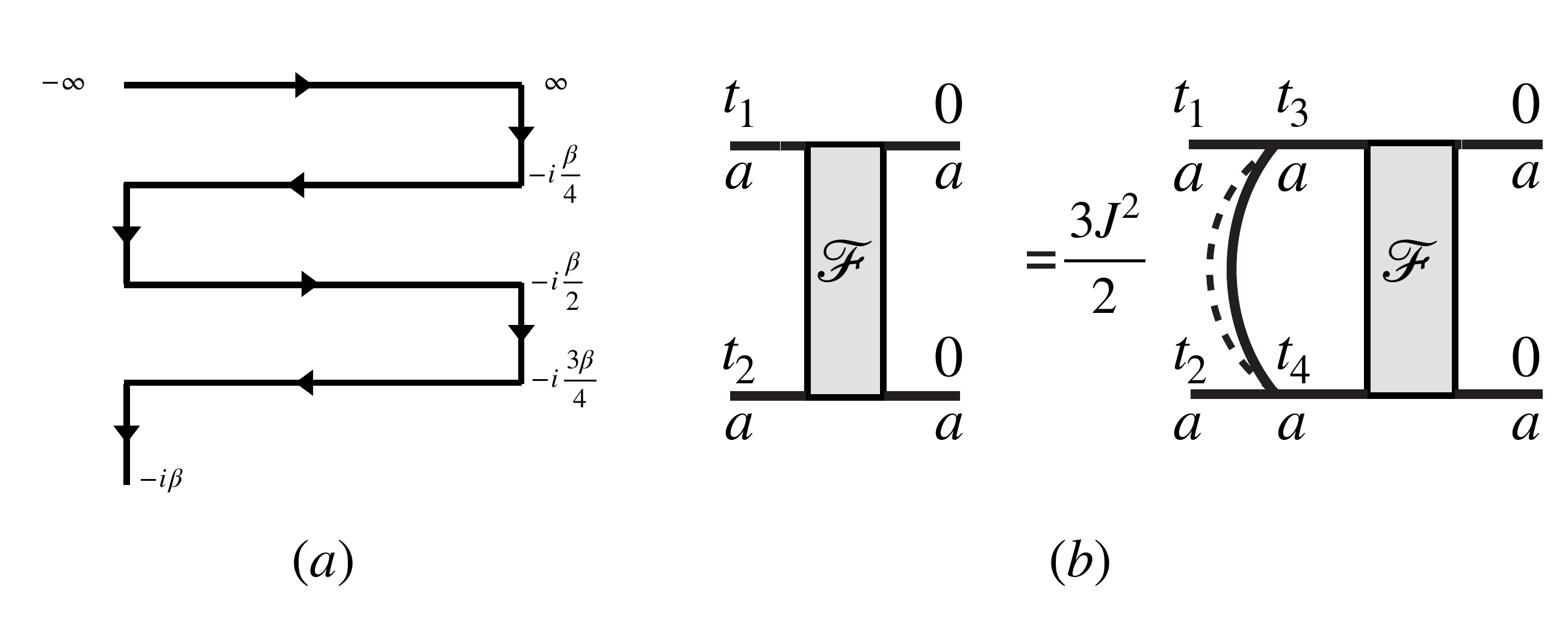}
	\caption{(a) The four real-time branches (separated by imaginary time $\beta/4$) of the Schwinger-Keldysh contour used for computing the OTOC.  (b) The ladder diagram for  Eq.\eqref{eq:KernelEq} for $\mathcal{O}(1/N)$ term ($\mathcal{F}_a$) in the OTOC $F_a(t_1, t_2)$ is shown for $p=3$. The solid horizontal lines denote dressed retarded propagator $Q^R_{aa}(t_1, t_3)$, $Q^R_{aa}(t_2,t_4)$ and the vertical rung denote Wightmann function $Q^W_{aa}(t_3, t_4)$. The dotted line represents disorder averaging. }\label{fig:LadderDiagram}
\end{figure} 

{\it OTOC and Lyapunov exponent.---}
As in the SYK model \cite{KitaevKITP,Maldacena2016}, the OTOC, $F(t)\sim \langle s_i(t)s_j(0)s_i(t)s_j(0)\rangle$ can be computed via real-time path integral method using a Schwinger-Keldysh (SK) contour with four branches, as shown in Fig.\ref{fig:LadderDiagram} \cite{KitaevKITP,Maldacena2016,BanerjeeAltman2016}. However, in contrast to the SYK model, where the large-$N$ saddle point is always replica symmetric \cite{Maldacena2016}, here we need to incorporate the non-trivial 1-RSB structure in the OTOC. We achieve this by using a replicated SK path integral \cite{Houghton1983,Cugliandolo2019} (SM, Sec. \textcolor{red}{S2}). We define the following regularized disorder-averaged OTOC \cite{Maldacena2015,Maldacena2016,Stanford2016}, $F_a(t_1,t_2)=(1/N^2)\sum_{ij}\mathrm{Tr}\left[ y s_{ia}(t_1)ys_{ja}(0)ys_{ia}(t_2)ys_{ja}(0)\right ]$, where $y^4=\exp{(-\beta H)}/\text{Tr}[\exp{(-\beta H})]$. The Lyapunov exponent $\lambda_\mathrm{L}$ is extracted from the chaotic growth, $\mathcal{F}_a(t,t)\sim e^{\lambda_\mathrm{L} t}$, that appears at $\mathcal{O}(1/N)$ in $F(t_1,t_2)$ (SM, Sec. \textcolor{red}{S2 1}). Over the intermediate-time chaos regime, $\lambda_\mathrm{L}^{-1}\lesssim t \lesssim \lambda_\mathrm{L}^{-1}\ln(N)$, $\mathcal{F}_a(t_1,t_2)$ can be obtained from a Bethe-Salpeter like equation \cite{KitaevKITP,Maldacena2016,BanerjeeAltman2016},
\begin{align}
\mathcal{F}_a(t_1, t_2) &= \int dt_3 dt_4 K_a(t_1,t_2,t_3,t_4)\mathcal{F}_a(t_3,t_4). \label{eq:KernelEq} 
\end{align}
The ladder Kernel $K$, e.g., for $p=3$, $K_a(t_1, t_2, t_3, t_4) = 3J^2Q^R_{aa}(t_{13})Q^R_{aa}(t_{24})Q^W_{aa}(t_{34})$ ($t_{13}=t_1-t_3$) (see SM, Sec. \textcolor{red}{S2 1}), is obtained using the retarded, $Q^R_{aa}(t)$, and the Wightmann, $Q^W_{aa}(t)=Q_{aa}(\tau\to \ci t+\beta/2)$, correlators \cite{Stanford2016}. For the chaotic growth regime, using the ansatz $F_a(t_1,t_2)=e^{\lambda_\mathrm{L}(t_1+t_2)/2}f_a(t_1-t_2)$ \cite{KitaevKITP,Maldacena2016,BanerjeeAltman2016}, $\lambda_\mathrm{L}$ is obtained by numerically diagonalizing the Kernel $K$ (SM, Sec. \textcolor{red}{S2 2}).

The information about the PM and SG phases are encoded in the ladder Kernel and a crucial difference is in the Wightmann correlator,
namely for the SG phase, $Q^W_{aa}(\w) = [2\pi \delta(\w)\qe - (\pi \rho(\w)/\sinh(\beta\w/2))]$,
whereas the first term is absent for $Q^W$ in the PM phase, where $q_\mathrm{EA}=0$. 

\begin{figure}[h!]
	\centering
	\includegraphics[width=1.0\linewidth]{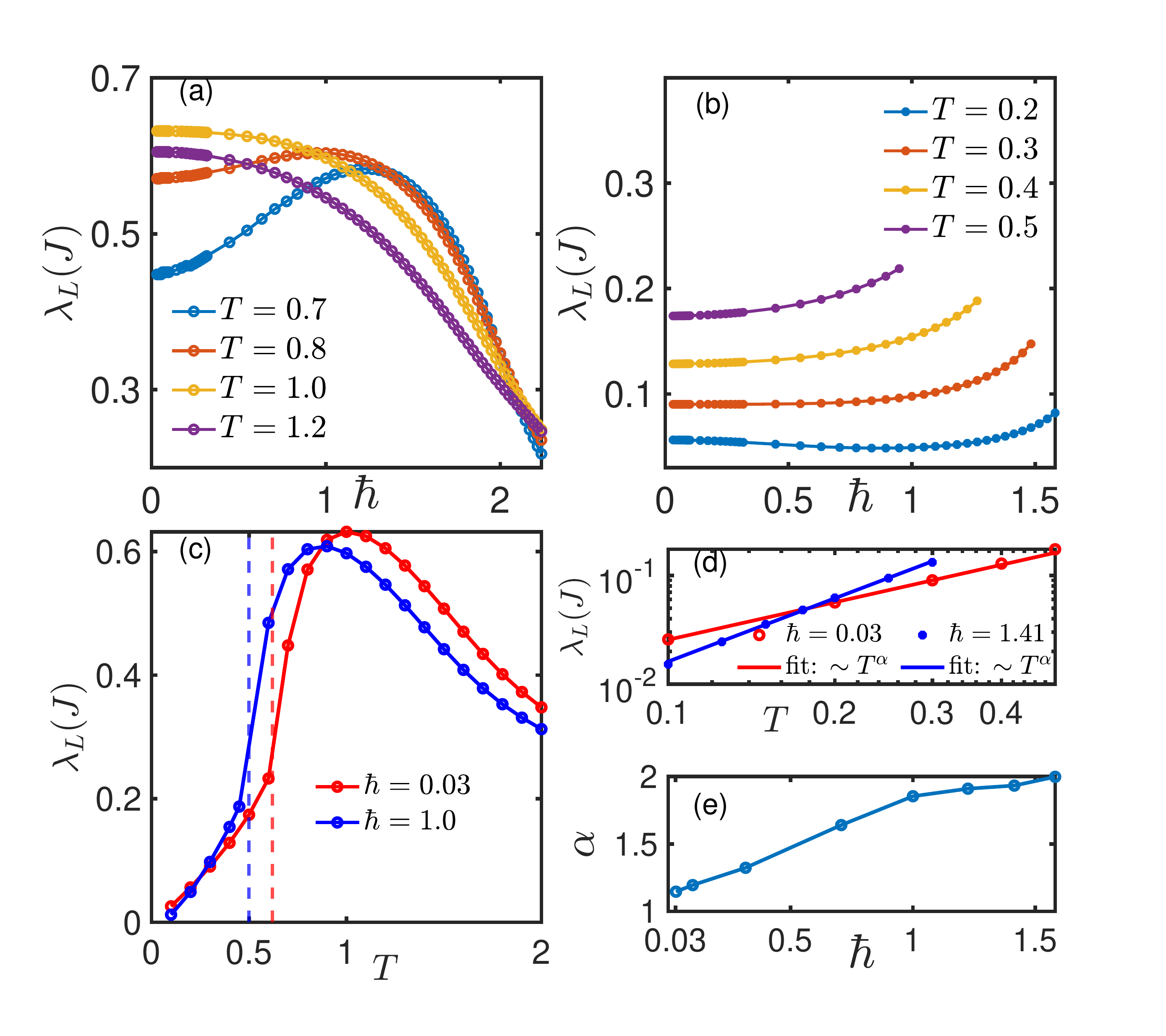}
	\caption{(a) Lyapunov exponent $\lambda_\mathrm{L}$ (in units of $J$) as a function of $\hbar$ for several values of $T$ in the cPM phase. 
	(b) $\lambda_\mathrm{L}$ (in units of $J$) as a function of $\hbar$ for several values of $T$ in the mSG  phase. 
	(c) $\lambda_L(T)$ (in units of $J$) across mSG-PM transitions ($T_d(\hbar)$, vertical dashed lines) for $\hbar=0.03, 1.0$.
	(d) $\lambda_\mathrm{L}$ vs. $T$ (log-log scale) at low temperature for $\hbar=0.03, 1.41$ fitted with a power-law  $\lambda_\mathrm{L}\propto T^{\alpha}$. (e) The exponent $\alpha$ varies from 1-2 as $\hbar$ varies from the classical limit to quantum limit.}\label{fig:lambdahbar}
\end{figure}

{\it Chaos in the paramagnetic phase.---}
We first discuss the dependence of $\lambda_\mathrm{L}$ on $T$ and $\hbar$ in the cPM phase, as shown through the colormap in Fig.\ref{fig:PhaseDiagram} for $p=3$. Overall, $\lambda_\mathrm{L}$ becomes small when $T$ or $\hbar$ are large. $\lambda_\mathrm{L}$ exhibits a broad maximum at $T_m(\hbar)$, substantially above $T_d$, albeit tracking $T_d(\hbar)$ line and merging with it at the tricritical point. $\lambda_\mathrm{L}$ is plotted in Fig.\ref{fig:lambdahbar}(a) as function of $\hbar$ for several temperatures. For  high and intermediate temperatures ($T\gtrsim J$), $\lambda_\mathrm{L}$ is a monotonically decreasing function of $\hbar$, approaching a constant value in the classical limit $\hbar\to 0$. $\lambda_\mathrm{L}$ decreases rapidly for $\hbar\gtrsim 1$ since the system acquires  a large spectral gap $\sim \Gamma$ for strong quantum fluctuations $\Gamma\gg T, J$ \cite{Cugliandolo2001} (SM, Sec. \textcolor{red}{S4 2}), making the interaction effects, and thus the chaos, very weak ($\lambda_\mathrm{L}\sim e^{-\Gamma/T}$). 

Remarkably, when temperature is close to $T_d(0)$, $\lambda_\mathrm{L}$ is a non-monotonic function of $\hbar$. This implies that the approach to classical limit
could be non-trivial for chaotic properties. 
A non-monotonic dependence is also seen with $T$ [Fig.\ref{fig:lambdahbar}(c)]. Starting from $T\gtrsim T_d(\hbar)$, $\lambda_\mathrm{L}$ initially increases reaching the maximum at $T_m$ and then decreases with increasing $T$, as $\sim 1/T^2$ at high temperature $T\gg J$ (SM, Sec. \textcolor{red}{S4 1}). In this limit, system has a small gap $\sim \sqrt{\Gamma T}< T$ for $T> \Gamma$,  whereas in the intermediate regimes $T,\Gamma\gtrsim T_m$, the system is soft-gappped and becomes gapless in the classical limit $\Gamma\to 0$ (SM, Sec. \textcolor{red}{S3}).

\begin{figure}[h!]
	\centering
	\includegraphics[width=1.0\linewidth]{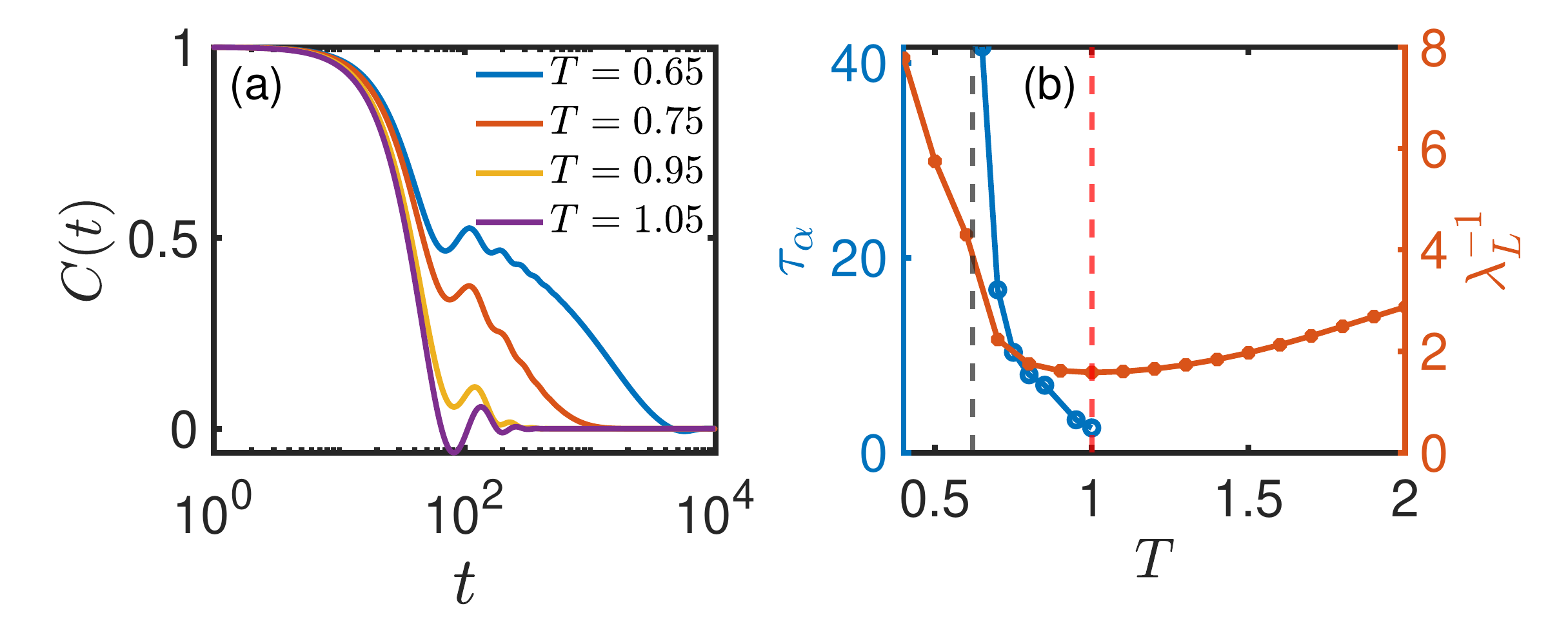}
 	\caption{(a) The correlation function $C(t)$ for several values of temperature $T$ in the classical limit ($\hbar=\sqrt{0.001}$).
 	(b) The $\alpha$-relaxation time scale $\tau_{\alpha}$, diverging for $T\to T_d$ (vertical black dashed line), extracted from $C(t)$ and $\lambda^{-1}_\mathrm{L}$ as function of $T$ in the classical limit $\hbar=\sqrt{0.001}$. 
The crossover temperature $T_\beta$ to two-step $\beta$-relaxation is shown by vertical red dashed line.}
	\label{fig:CorrelationFn}
\end{figure}
{\it Chaos in the spin glass phase.---} In contrast to the PM phases, the mSG phase is gapless \cite{Cugliandolo2000,Cugliandolo2001}. Moreover, unlike that in the cPM phase [Fig.\ref{fig:lambdahbar}(a)], $\lambda_\mathrm{L}(\hbar)$, in general, monotonically increases with $\hbar$, as shown in Fig.\ref{fig:lambdahbar}(b), apart from some weak non-monotonic dependence on $\hbar$ at low temperatures. Thus, quantum fluctuations makes the system more chaotic in the mSG phase. Fig.\ref{fig:lambdahbar}(c) shows $\lambda_\mathrm{L}(T)$ for two $\hbar$ values. The Lyapunov exponent follows a power-law temperature dependence, $\lambda_\mathrm{L}\sim T^\alpha$, with exponent $\alpha$ varying from 2 to 1 [Fig.\ref{fig:lambdahbar}(d, e)] with decreasing $\hbar$, implying $\lambda_\mathrm{L}\sim T$ in the classical limit. However, the pre-factor of linear $T$ is much smaller than $2\pi/\hbar$ corresponding to the bound (see SM, Sec. \textcolor{red}{S4 3}).

The temperature dependence $\lambda_\mathrm{L}\sim T^2$, for large $\hbar$ [Fig.\ref{fig:PhaseDiagram}] within the mSG phase, is similar to that in a Fermi liquid \cite{BanerjeeAltman2016,Kim2020,Kim2020a}. This $T$-dependence in the mSG phase can be understood based on the observation that the self-consistent equations for the time-dependent part of $Q_{ab}(\tau)$ [Eq.\eqref{eq:SaddlePoint}], and the Kernel [Eq.\eqref{eq:KernelEq}], in the presence of 1-RSB are equivalent to those in the PM phase of an effective model with both $p=3$ ($J_3=J$) and $p=2$ ($J_2=J\sqrt{3q_{EA}}$) terms in Eq.\eqref{eq:Hamiltonian} (SM, Sec. \textcolor{red}{S2 1}). Irrespective of the $J_3/J_2$ ratio, $p=3$ term is irrelevant at low energy and can be treated perturbatively, with a Lagrange multiplier $z=2J_2$ such that the system is gapless like in the mSG. In this case, as discussed in the SM, Sec. \textcolor{red}{S4 4}, the integral Kernel equation in Eq.\eqref{eq:KernelEq} can be converted into,
\begin{align*}
\left(-\frac{1}{2}\frac{\partial^2}{\partial t^2}-\sech^2{t}\right)f(t)={-}\frac{1}{3}\left(\lambda_\mathrm{L}\frac{J_2^3}{\pi J_3^2 T^2}+1\right)f(t),
\end{align*}
i.e. a one dimensional Schr\"{o}dinger equation with P\"{o}schl-Teller potential, with well-known eigenvalues \cite{Poschl1933}. This leads to $\lambda_\mathrm{L}\sim T^2/q_\mr{EA}^{3/2}$. The exponent $\alpha\simeq 2$ matches with numerically obtained value in Fig.\ref{fig:lambdahbar}(e) in the quantum limit for large $\hbar$, where $J_2\propto \sqrt{q_{EA}}$ is weakly temperature dependent (SM, Sec. \textcolor{red}{S4 4}). Since quantum fluctuations reduces the SG order parameter $q_\mr{EA}$ (SM, Sec. \textcolor{red}{S4 4}), $\lambda_\mr{L}\sim q_\mr{EA}^{-3/2}$ naturally explains the enhancement of chaos [Fig.\ref{fig:lambdahbar}(b)] due to $\hbar$.



{\it Scrambling from glassy relaxation.---} As shown in Fig.\ref{fig:CorrelationFn} (a), the decay of spin-spin correlation $C(t)=(1/N)\sum_i\langle s_i(t)s_i(0)$, becomes slower as $T\to T_d(\hbar)^+$. Moreover, close to $T_d$, $C(t)$ exhibits a two-step relaxation, typical characteristic of supercooled liquids \cite{GotzeBook,Kob1997,Reichman2005}, namely -- (1) a fast microscopic decay followed by a slowly decaying plateau-like $\beta$-relaxation regime, and eventually (2) the $\alpha$ regime with a stretched exponential decay $\sim \exp{[-(t/\tau_\alpha)^{\beta_a}]}$, with a diverging time scale $\tau_\alpha \sim (T-T_d)^{-\gamma}$ ($\gamma>0$) and stretching exponent $\beta_a$ \cite{GotzeBook,Kob1997,Reichman2005}. The emergence of the two-step relaxation close to $T_d$ is seen in Fig.\ref{fig:CorrelationFn}(a) (SM, Sec. \textcolor{red}{S5}).
In Fig.\ref{fig:CorrelationFn}(b), we plot $\tau_\alpha$ extracted from the numerical fit to $C(t)$ [Fig.\ref{fig:CorrelationFn}(a)] in the $\alpha$ regime and compare with $\lambda_\mathrm{L}^{-1}$. In contrast to $\tau_\alpha$, $\lambda_\mathrm{L}^{-1}$ has a minimum at $T_m$, substantially above $T_d$. 
In \textcolor{red}{SM, Sec. S6 1}, we show that the stretched exponential part from $\alpha$-relaxation of $C(t)$ [Fig.\ref{fig:CorrelationFn}(a)] alone with the $\tau_\alpha$ shown in Fig.\ref{fig:CorrelationFn}(b) give rise to the non-monotonic $\lambda_\mathrm{L}(T)$ 
[Fig.\ref{fig:lambdahbar}(c)]. 
We analytically solve the Kernel equation [Eq.\eqref{eq:KernelEq}] and obtain $\lambda_\mathrm{L}(T)$ in the PM phase for Debye relaxation  $\sim \exp(-t/\tau_\alpha)$ \textcolor{red}{(SM, Sec. S6 2)}. We show that 
\begin{align}
\lambda_\mathrm{L}&\sim \tau_\alpha^{-1}(2J/T-1)\nonumber
\end{align}
for $T\lesssim J$, leading to a maximum at $T_m\sim \sqrt{JT_d}$ for $\gamma=1$. Thus, for temperature close to $T_d$, $\lambda_\mathrm{L}\propto \tau_\alpha^{-1}$, i.e. the scrambling rate is controlled by relaxation rate. However, the crossover from strong ($J>T$ ) to weak ($J<T$) coupling in combination with the non-trivial temperature dependence of $\tau_\alpha$ in the glassy regime, give rise to a crossover in chaos in the form of a maximum in the Lyapunov exponent. As shown in Fig.\ref{fig:PhaseDiagram}, we find that $T_m$ is correlated with the crossover ($T_\beta$) to two-step glassy relaxation [Fig. \ref{fig:CorrelationFn}(b)]. The onset of non-trivial temperature dependence of $\tau_\alpha$ is presumably connected with the onset of the two-step relaxation, leading to the correlation between $T_m$ and $T_\beta$. However, to properly establish this relation, we need an analytical understanding of $C(t)$ and $T_\beta$, which is beyond the scope of this letter.

{\it Conclusions.---} In this work, we have shown how  quantum mechanics influences chaos in a solvable quantum spin glass model.
We derive relation between chaos and relaxation rates in the complex glassy regime above the glass transition. So far, such direct relation between scrambling and relaxation has only been established for weakly interacting  systems \cite{Aleiner2016,Stanford2016,BanerjeeAltman2016,Patel2017}. In future, studying the connection between many-body chaos and relaxation in simulation \cite{Sastry1996} of super-cooled liquids \cite{Berthier2011} may lead to new insight into complex dynamics in glasses. It would be interesting to study the quantum to classical crossover in other models \cite{Scaffidi2019}, where $\lambda_\mathrm{L}$ already starts at the upper bound $2\pi T/\hbar$ in the quantum limit. Also, the methods developed here to analyze chaos can be extended to transverse-field models \cite{Goldschmidt1990,Dobrosavljevic1990,Nieuwenhuizen1998,Obuchi2007,Nieuwenhuizen1998}, e.g. with Ising spins. 

\emph{Note added.--} After this work came out as an arXiv preprint, Ref.\onlinecite{Anous2021}, which looks into quantum $p$-spin glass model and its chaotic properties from holographic perspective, appeared on the arXiv.

VL acknowledges support from CSIR, India. SB acknowledges support from SERB (ECR/2018/001742), DST, India.

\bibliography{CrossoverQuantumToClassical}

\def\makeSM{1}
\ifdefined\makeSM

\appendix
\renewcommand{\appendixname}{}
\renewcommand{\thesection}{{S\arabic{section}}}
\renewcommand{\theequation}{\thesection.\arabic{equation}}
\renewcommand{\thefigure}{S\arabic{figure}}
\setcounter{page}{1}
\setcounter{figure}{0}
\setcounter{equation}{0}

\widetext

\centerline{\bf Supplemental Material}
\centerline{\bf for}
\centerline{\bf \titlename}
\centerline{by \authornames}
\affiliation{\affiliations}
\fi 

\def  \qinv{Q^{-1}}
\def  \q0{\frac{\w_k^2}{\G}+z}

\newcommand{\lin}{linspace}
\newcommand{\m}{\Delta_0}
\newcommand{\M}{\Delta}
\newcommand{\g}{Q}
\newcommand{\ginv}{\big(Q^{-1}\big)_{ab}}
\newcommand{\qr}{q_{\text{reg}}} 
\newcommand{\sr}{\Sigma_{\text{reg}}}
\newcommand{\qt}{\tilde{q}_{{EA}}}
\newcommand{\bfig}{\begin{figure}[H]\centering}
\newcommand{\efig}{\end{figure}}
\newcommand{\s}{\hspace{0.5cm }} 
\newcommand{\sh}{\hspace{0.25cm }} 

\section{Imaginary-time path integral and saddle point equations}\label{suppSD}

The imaginary time path-integral and saddle point equations for $p$-spin glass model are discussed in Refs.\onlinecite{Cugliandolo2000,Cugliandolo2001}. Here we sketch the basic steps of the calculations. The thermodynamic properties of the model [Eq.\eqref{eq:Hamiltonian}] is obtained from the free-energy,
\begin{align}\label{S04}
	F = -k_B T\overline{\ln Z},
\end{align}
where the overline denotes disorder average over  configuration $\{J^{(p)}_{i_1\dots i_p} \}$ of the random $p$-spin couplings and $T$ is the temperature of the system. We use the replica trick, $\overline{\ln Z} =\lim_{n\rightarrow 0} (\overline{Z^n}-1)/n$, to obtain disorder averaged free energy. The replicated imaginary-time path integral is obtained as

\begin{align}
Z^n  &= \int \left[\prod_a \mathcal{D}s_a(\tau)\delta \left(\sum_{i}s^2_{i,a}(\tau)-N\right)\right]\exp\bigg[-\frac{1}{\hbar}\int^{\beta\hbar}_0 d\tau \bigg(\sum_{i,a}\frac{M}{2}\left(\frac{\partial s_{i,a}}{\partial \tau }\right)^2 + \sum_{p,i_1<...<i_p, a}J^{(p)}_{i_1...i_p}s_{i_1,a}(\tau)...s_{i_p,a}(\tau)\bigg) \bigg]
\end{align}
where $a=1,\dots,n$ is the replica index and the $\delta$-function in the above imposes the spherical constraint. We write the $\delta$-function as $\int  \mathcal{D}z_a(\tau) \exp\big[-\int_{0}^{\beta\hbar} z_a(\tau)\big(\sum_{i,a}s^2_{i,a}(\tau)-N\big)\big]$.
After averaging over disorder, introducing bi-local order parameter field $Q_{ab}(\tau,\tau')=(1/N)\sum_{i} s_{i,a}(\tau)s_{i,b}(\tau')$ and integrating out the fields $\{s_{ia}(\tau)\}$, we obtain an effective action in terms of $Q_{ab}(\tau,\tau')$, i.e.
\begin{align} \label{S1}
\overline{Z^n} &= \int \mathcal{D}Q_{ab}\exp\big[-{S_{\text{eff}}[Q_{ab}]}/{\hbar}\big].
\end{align}
For the equilibrium saddle point $Q_{ab}(\tau, \tau')=Q_{ab}(\tau-\tau')$ and  we use Matsubara frequencies, $\omega_k=2\pi kT$ with integers $k$, to write down the effective action as 
\begin{align} \label{Seff}
-\frac{S_{\text{eff}}}{\hbar} &= \frac{N}{2}\sum_{k}\Tr \ln[(\beta\hbar)^{-1}\boldsymbol{Q}(\w_k)] + \frac{N}{2}\sum_{k, a, b}\bigg[\delta_{ab}-\frac{1}{\hbar}( M {\w^2_k}+z)\delta_{ab}Q_{ab}(\w_k) \bigg] \notag \\
& + \frac{J^2 N \beta}{4\hbar}\sum_{a, b} \int_{0}^{\beta\hbar} d\tau \bigg[\frac{1}{\beta\hbar}\sum_{k}\exp\big(-\ci\w_k\tau\big)Q_{ab}(\w_k)\bigg]^p + \frac{Nn\beta}{2}z
\end{align}
where $z_a(\tau)=z$ and $Q_{ab}(\w_k) = \int_{0}^{\beta\hbar} d\tau \exp(\ci\w_k\tau)Q_{ab}(\tau)$.  
For $N\rightarrow \infty$ limit the following saddle-point equation is obtained by minimizing the above effective action as
\begin{align} \label{sg:1}
\frac{1}{\hbar}\big(M\w^2_k + z\big)\delta_{ab} &= \ginv (\w_k) + \sum_p\frac{J^2_p p}{2\hbar^2}\int_{0}^{\beta\hbar} d\tau \exp(\ci\w_k\tau)[Q_{ab}(\tau)]^{p-1}
\end{align}
 We work with the dimensionless quantities - $\tilde{J_p}=J_p/J$, energy and temperature in units of $J$, and time $\tilde{\tau}=J \tau/\hbar$ and frequency $\tilde{\w_k}=\w_k\hbar/J$, to obtain
 {
	\begin{align} \label{sg:2}
	\ginv(\tilde{\w}_k) &= Q^{-1}_{0}(\tilde{\w}_k)\delta_{ab} - \Sigma_{ab}(\tilde{\w}_k)
	\end{align}
}    
where 
\begin{align}
Q^{-1}_{0}(\tilde{\w}_k) &= \frac{\tilde{\w}^2_k}{\Gamma}+\tilde{z},  \hspace{0.5cm} \Gamma = \hbar ^2/MJ \notag \\ \label{sg:3}
\Sigma_{ab}(\tilde{\w}_k) &= \sum_p\frac{\tilde{J}_p^2 p}{2}\int_{0}^{\beta} d\tilde{\tau} \exp(\ci\tilde{\w}_k\tilde{\tau})[Q_{ab}(\tilde{\tau})]^{p-1}
\end{align}
with $\tilde{z}=z/J$. The quantum fluctuation parameter $\hbar$ enters through the parameter $\Gamma=\hbar^2/MJ$ in the saddle point equation.  
From here on, we work with these dimensionless variables and, to simplify the notation, we omit the hat from the symbols.

\subsection{Saddle point equation and spectral function calculation in the PM phase}\label{supp-SD-PM}

In the PM phase, the order parameter is replica symmetric and diagonal, i.e. $Q_{ab}(\w_k)=Q(\w_k)\delta_{ab}$. So, the saddle point Eqs.\eqref{sg:2} and \eqref{sg:3} in the PM phase, e.g., for $p=3$ spin glass model simplify to
\begin{subequations}\label{eq:PM-im-SD}
\begin{align}
\qinv (\w_k) = \frac{\w_k^2}{\G} + z - \Sigma(\w_k) \\
\Sigma(\tau) = \frac{3 \tilde{J_3}^2}{2}[Q(\tau)]^2
\end{align}
\end{subequations}
To obtain the spectral function, we solve the above equations numerically after analytical continuation from Matsubara to real frequency i.e  $\ci\w_k\rightarrow \w+\ci 0^+ $, i.e. in terms of the retarded functions
\beq \label{eq:PM-SD}
\left( Q^R (\w)\right)^{-1} = -\frac{{\w}^2}{\G} + z - \Sigma^R(\w)
\eeq 
where $\Sigma_R(\w)$ is obtained from $\Sigma(\w_k)$, i.e.
\beq 
\Sigma(\w_k) = \int_{0}^{\beta} d\tau e^{\ci \w_k \tau} \Sigma(\tau) = 
\frac{3\tilde{J_3}^2}{2\beta}\sum_{k} Q(\w_k)Q(\w_n-\w_k)
\eeq 
We perform the Matsubara summation using the spectral representation $Q(\w_k)=\int_{-\infty}^{\infty} d\w \rho(\w)/(\ci\w_k-\w)$, where $\rho(\w)=-\text{Im}Q^R(\w)/\pi$ is the spectral function. After analytical continuation, we obtain 
\beq 
 \Sigma^R(\omega) = \frac{3\tilde{J}_3^2}{2}\int_{-\infty}^{\infty} \int_{-\infty}^\infty d\omega_1 d\omega_2 \rho(\omega_1)\rho(\omega_2)\frac{n_\mathrm{B}(-\omega_1)n_\mathrm{B}(-\omega_2)-n_\mathrm{B}(\omega_1)n_\mathrm{B}(\omega_2)}{\omega_1+\omega_2-\omega-\ci 0^+}.
\eeq 
Here $n_\mathrm{B}(\omega)=1/(e^{\beta \omega}-1)$ is the Bose function. To numerically evaluate the self-energy $\Sigma^R(\w)$ efficiently, we use the identity $1/(\w-\w_1-\w_2+\ci 0^+)=-\ci \int_{0}^{\infty}dt e^{\ci(\w-\w_1-\w_2+\ci 0^+)t}$ to rewrite the above equation as follows
 \beq\label{pm:sigmaW}
 \Sigma^R(\omega) = \ci \tilde{J_3}^2 \int_{0}^{\infty} dt e^{\ci \omega t} [n_1(t)^2 - n_2(t)^2],
 \eeq
 where 
 $
 n_1(t) = \int_{-\infty}^{\infty} d\omega e^{-\ci \omega t}\rho(\omega)n_B(-\omega)
 $
 and 
 $
 n_2(t) = \int_{-\infty}^{\infty} d\omega e^{-\ci \omega t}\rho(\omega)n_B(\omega)
 $. 
 
The Lagrange multiplier $z$ in Eq.\eqref{eq:PM-SD} can be determined either from the imaginary-time calculation using Eqs.\eqref{eq:PM-im-SD} subjected to the spherical constraint $Q_{aa}(\tau=0)=1$, or we can determine $z$ from  real frequency calculation itself. To determine $z$ from real frequency calculation, we can vary $z$ and the correct $z$ will give physical solution which satisfies the spherical constraint expressed in terms of the spectral function, 
\beq \label{eq:PM-18}
-\int_{-\infty}^\infty d\w \rho(\w)n_B(\w) = 1.
\eeq 
We use the above sum rule condition Eq.\eqref{eq:PM-18} to check the accuracy of the spectral function obtained numerically by iterating the saddle-point equations.
 
We solve the self-consistent saddle point equations \eqref{eq:PM-SD}, \eqref{pm:sigmaW} by discretizing over frequency ($\w$) and starting with some initial guesses for Lagrange multiplier $z$ and $Q^R(\w)$, e.g. the non-interacting retarded function for $J_3=0$.
 We calculate the retarded self-energy $\Sigma^R(\w)$ and $n_i(t)$, $i=1, 2$ using fast-Fourier transform (FFT) and iterate for $Q^R(\w)$ until convergence with a required numerical accuracy. We repeat this process by varying $z$ till the sum rule condition of Eq.\eqref{eq:PM-18} is satisfied. We find that the process is much more efficient if we split $z$ as $z=z'+\Sigma^R(\w=0)$ and vary $z'$.

\subsection{Saddle point equation in the SG phase}\label{supp-SD-SG}
In the quantum spherical $p$-spin glass model, there are two types of SG phases -- a thermodynamic SG phase and a marginal SG phase. A detailed discussion can be found in Refs.\onlinecite{Cugliandolo2000,Cugliandolo2001}. Here we briefly discuss these two different SG phases for completeness and then mainly focus on the saddle-point solution for the marginal SG phase. 

 The replica-symmetric spin-glass phase, unlike that for $p=2$ \cite{Kosterlitz1976}, is unstable for $p\geq 3$ and one needs to consider the replica symmetry breaking. In this model, one step replica symmetry (1-RSB) breaking is exact \cite{Cugliandolo2001} and the order parameter in the imaginary time is given by 
\beq \label{sp:Qabtau}
Q_{ab}(\tau) = (q_d(\tau)-\qe)\delta_{ab} + \qe \epsilon_{ab}
\eeq 
where $\qe$ is Edward-Anderson order parameter and $\epsilon_{ab}=1$ for the diagonal blocks and zero otherwise, as described in the main text also. The above 1-RSB order parameter in Matsubara frequency reads as 
	\beq \label{rsb1}
	Q_{ab}(\w_k) = (\qd(\w_k)-\qt)\delta_{ab}+\qt\epsilon_{ab},
	\eeq
	where $\qt=\beta \qe \delta_{\w_k, 0}$ and $\qd(\w_k)$ is Matsubara Fourier transformation of $q_d(\tau)=Q_{aa}(\tau)$. Now, it is convenient to write $q_d(\tau) = \qr(\tau) + \qe$.
The inverse matrix $\boldsymbol{Q^{-1}}(\w_k)$ has the following structure 
\beq \label{rsb2}
\ginv(\w_k) = A(\w_k)\delta_{ab} + B(\w_k)\epsilon_{ab}
\eeq 
with
\beq \label{rsb3}
A(\w_k) = \frac{1}{\qd(\w_k)-\qt}, \hspace{0.8cm}
B(\w_k) = \frac{-\qt}{\qd(\w_k)^2-(m-1)\qt^2+(m-2)\qd(\w_k)\qt}.
\eeq  
Here $m$ is the break point\cite{Cugliandolo2001}.   
The diagonal element of the inverse matrix ($\boldsymbol{Q^{-1}}$) is given by
\beq \label{rsb4}
(Q^{-1})_{aa} = A(\w_k) + B(\w_k) = \frac{\qd(\w_k)+(m-2)\qt}{(\qd(\w_k)^2+(m-2)\qt\qd(\w_k)-(m-1)\qt^2)}.
\eeq      
The off-diagonal element ($a\neq b$) is 
\beq \label{rsb5}
\ginv(\w_k) = B(\w_k).
\eeq   
The saddle-point equation for diagonal component $\qd(\w_k)$ can be obtained by using equations \eqref{sg:2} and \eqref{rsb4} as 
	\beq \label{rsb6}
	\frac{\w_k^2}{\G}+z = \frac{\qd(\w_k)+(m-2)\qt}{(\qd(\w_k)^2+(m-2)\qt\qd(\w_k)-(m-1)\qt^2)} + \Sigma(\w_k),
	\eeq 
with $\Sigma(\w_k)=\Sigma_{aa}(\w_k)$, the Fourier transform of $\Sigma_{aa}(\tau)=\frac{\tilde{J}_p^2p}{2}[q_d(\tau)]^{p-1}$. 
 Using equations \eqref{sg:2} and \eqref{rsb5}, the saddle point equation for $\qe$ is obtained from
{
	\beq \label{rsb7}
	0 =-\frac{\qe}{\qd(0)^2-(m-1)\beta^2\qe^2+(m-2)\beta\qd(0)\qe} + \frac{ p}{2} \qe^{p-1}.
	\eeq }
It is convenient to separate out constant and $\tau-$dependent part of $q_d(\tau)$ and self-energy $\Sigma_{aa}(\tau)$. So, we write these as below 
\beq \label{sg:t-const}
\qr(\tau) = q_d(\tau)-\qe, \hspace{0.7cm}  \sr(\tau)= \frac{\tilde{J}_p^2p}{2}[q_d(\tau)]^{p-1} -\frac{\tilde{J}_p^2p}{2}\qe^{p-1}.
\eeq 
 Using saddle point Eq.\eqref{rsb6}, we can rewrite Eq. \eqref{rsb6} as follows
\beq 
\frac{\w^2_k}{\G}+z = \frac{1}{\qr(\w_k)}+\sr(\w_k)
+ \bigg[ \frac{-\beta\qe\delta_{\w_k, 0}}{\qd(\w_k)^2-(m-1)\qe^2\beta^2\delta_{\w_k, 0}+(m-2)\qd(\w_k)\qe\beta\delta_{\w_k,0}} + \frac{p}{2}\qe^{p-1}\beta\qe\delta_{\w_k, 0}\bigg] \notag \\
\eeq 
The term inside the third bracket above is zero due to Eq.\eqref{rsb7}, and we obtain the simplified saddle point equation
	\beq \label{sg:sd1}
	\frac{\w_k^2}{\G} + z = \frac{1}{\qr(\w_k)} + \sr(\w_k),
	\eeq 
	where 
	\beq \label{sg:sd-sigma}
		\sr(\w_k) = \frac{\tilde{J}^2_p p}{2}\int_{0}^{\beta} d\tau e^{\ci\w_k\tau}\times (q_d^{p-1}(\tau)-\qe^{p-1}).
  \eeq 
 Moreover, it is convenient to split the Lagrange multiplier $z$ as $z=z'+\sr(\w_k=0)$ with
  	\beq 
  	z' \sh=\sh \frac{1}{\qd(0)(1-y)} 
  	= \frac{p}{2}\beta m \qe^{p-1} \frac{1+x_p}{x_p},
  	\eeq 
	where $y={\beta\qe}/{\qd(0)}$ and $x_p={my}/{(1-y)}$. The saddle point equation [Eq.\eqref{rsb6}] for the Edward-Anderson parameter $\qe$ can be rewritten as 
	\beq \label{sg:qe}
	\frac{p(\beta m)^2}{2}\qe^p = \frac{x_p^2}{1+x_p}.
	\eeq
	
	In the thermodynamic SG phase, the break point $m$ is determined by extremizing the free-energy functional $F[m]$ \cite{Cugliandolo2001}, leading to 
	\beq 
	\ln\bigg[\frac{1}{1+x_p} \bigg] + \frac{x_p}{1+x_p} + \frac{x_p^2}{p(1+x_p)}    = 0.
	\eeq 
	
	On the other hand, in the marginal SG phase, the marginal stability condition\cite{Cugliandolo2001} is applied, i.e the replicon eigenvalue or the transverse eigenvalue of gaussian fluctuation matrix around the 1-RSB saddle point is set to zero to determine the break point $m$. The replicon eigenvalue $\Lambda_T$ is given by\cite{Cugliandolo2001} 
	\beq 
	\Lambda_T = \beta^2\bigg[ \bigg(\frac{1}{\qd(0)-\beta\qe}\bigg)^2-\frac{p(p-1)}{2}\qe^{p-2}\bigg]
	\eeq 
	Setting $\Lambda_T=0$, we obtain $x_p=p-2$, which determines $m$ in marginal SG phase.
	
\subsection{Numerical solution of the saddle point equation in the marginal SG phase}\label{supp-SD-SG-sol}
To solve the saddle point equation for the marginal SG phase, e.g. for $p=3$, 
using $q_d(\tau)=\qr(\tau)+\qe$, we rewrite Eq.\eqref{sg:sd-sigma} as
\beq
\sr(\w_k) = \Sigma_3(\w_k) + 3{J}^2\qe \qr(\w_k), \hspace{0.7cm}\Sigma_3(\w_k)=\frac{3{J}^2}{2}\int_{0}^{\beta}d\tau e^{\ci\w_k\tau}[\qr(\tau)]^2 
\eeq 
Therefore, the self-energy $\sr(\w_k)=\Sigma_3(\w_k) + \Sigma_2(\w_k)$, and thus the saddle point equation, formally resemble those in the PM phase of the model [Eq.\eqref{eq:Hamiltonian}] with both $p=2$ and $p=3$ terms with $J_3=J$ and $J_2=\sqrt{3\qe} J$. 

Now, to obtain the spectral function we analytical continue Eq.\eqref{sg:sd1} to real frequency to get
\beq 
({\qr^R})^{-1}(\w) = -\frac{\w^2}{\Gamma}+z - \Sigma^R_3(\w)-3J^2\qe\qr^R(\w).
\eeq 
The spectral function in this phase is defined as
\beq 
\rho(\w)  = -\frac{1}{\pi}\text{Im}[\qr^R(\w)]
\eeq 
So, we can express $\qr(\tau)$ in terms spectral representation as 
\beq \label{sp:qrspectral}
\qr(\tau) = -\int_{-\infty}^\infty d\w \rho(\w)n_B(\w)e^{\w(\beta- \tau)}
\eeq 
The spherical constraint $Q_{aa}(\tau=0)=q_d(\tau=0)=1$ in terms of spectral function is obtained as
\beq 
 -\int_{-\infty}^\infty d\w \rho(\w)n_B(\w)  = 1-\qe.
\eeq 
So, this is a sum rule condition for spectral function in the mSG phase. 
Similar to the PM case, using spectral representation we can rewrite $\Sigma^R_3(\w)$ in the form of Eq.\ref{pm:sigmaW}. To determine $\qe$, we use Eq.\ref{sg:qe} and marginal stability criterion $x_p=1$, e.g., for $p=3$. We follow similar steps, as discussed for the PM phase in Sec.\ref{supp-SD-PM}, to obtain numerically converged solution for which the above sum rule condition is satisfied.

\section{Real-time Schwinger-Keldysh contour for OTOC for paramagnetic and spin glass phase}\label{supp-SK-form}
Here we briefly discuss the Schwinger-Keldysh (SK) formulation for the disorder averaged regularized OTOC \cite{Maldacena2015},
\begin{align}
F(t_1,t_2)=\frac{1}{N^2}\sum_{ij} \overline{\mathrm{Tr}[ys_i(t_1)ys_j(0)ys_i(t_2)ys_j(0)]}, \label{eq:OTOCop}
\end{align}
where $y^4=\exp(-\beta H)/\mathrm{Tr}[\exp(-\beta H)]$. The above correlation function can be formulated as a many-body path integral \cite{KamenevBook} on a Schwinger-Keldysh (SK) contour with four real-time branches that are separated by $\beta/4$ in the imaginary-time \cite{Maldacena2016,Stanford2016,BanerjeeAltman2016}, namely by writing the path integral representation for the following generating function
\begin{align}
\mathcal{Z}=\frac{1}{Z}\mathrm{Tr}\left[e^{-\beta H/4}U(t_0,t_f)e^{-\beta H/4}U(t_f,t_0)e^{-\beta H/4}U(t_0,t_f)e^{-\beta H/4}U(t_f,t_0)\right]. \label{eq:SKGeneratingF}
\end{align}
Here $Z=\mathrm{Tr}[\exp(-\beta H)]$ is the equilibrium partition function and $U(t,t_0)=\mathcal{T}_t\exp[-(\ci/\hbar)\int_{t_0}^tH(t')dt']$ is the time evolution operator with $\mathcal{T}$ denoting time ordering; $t_0$ is the initial and $t_f$ some final time. In the Keldysh formalism, typically the disorder averaging can be performed without introducing replicas \cite{KamenevBook}. This procedure can be used to describe the real-time dynamics and compute dynamical correlations such as OTOC, e.g., in the PM phase of the quantum $p$-spin glass model. However, to obtain the dynamical correlations in the SG phase, one needs to consider the actual physical situation, e.g., the initial condition or the initial density matrix more carefully \cite{Houghton1983,Cugliandolo2019}. For instance, to have a clear physical situation in mind, one can weakly couple the system at $t_0$ with a bath at temperature $T$, while the systems is at a higher temperature $>T$, and let the system cool down to the temperature $T$ \cite{Cugliandolo2001} in the presence of bath. In this case the system could be equilibrated up to $T\gtrsim T_d$ in the PM phase. Hence, to compute the OTOC in the PM phase  above $T_d$, we can directly disorder average the generating function $\mathcal{Z}$ [Eq.\eqref{eq:SKGeneratingF}] and take $t_0\to -\infty$ and $t_f\to \infty$ limits and vanishing coupling with the bath. In this way, the correlation and retarded functions become time-translational invariant and identical to the equilibrium ones obtained by solving the saddle-point equations in the PM phase in Sec.\ref{supp-SD-PM}.

\begin{figure}[htb]
	\centering
	\includegraphics[scale=0.3]{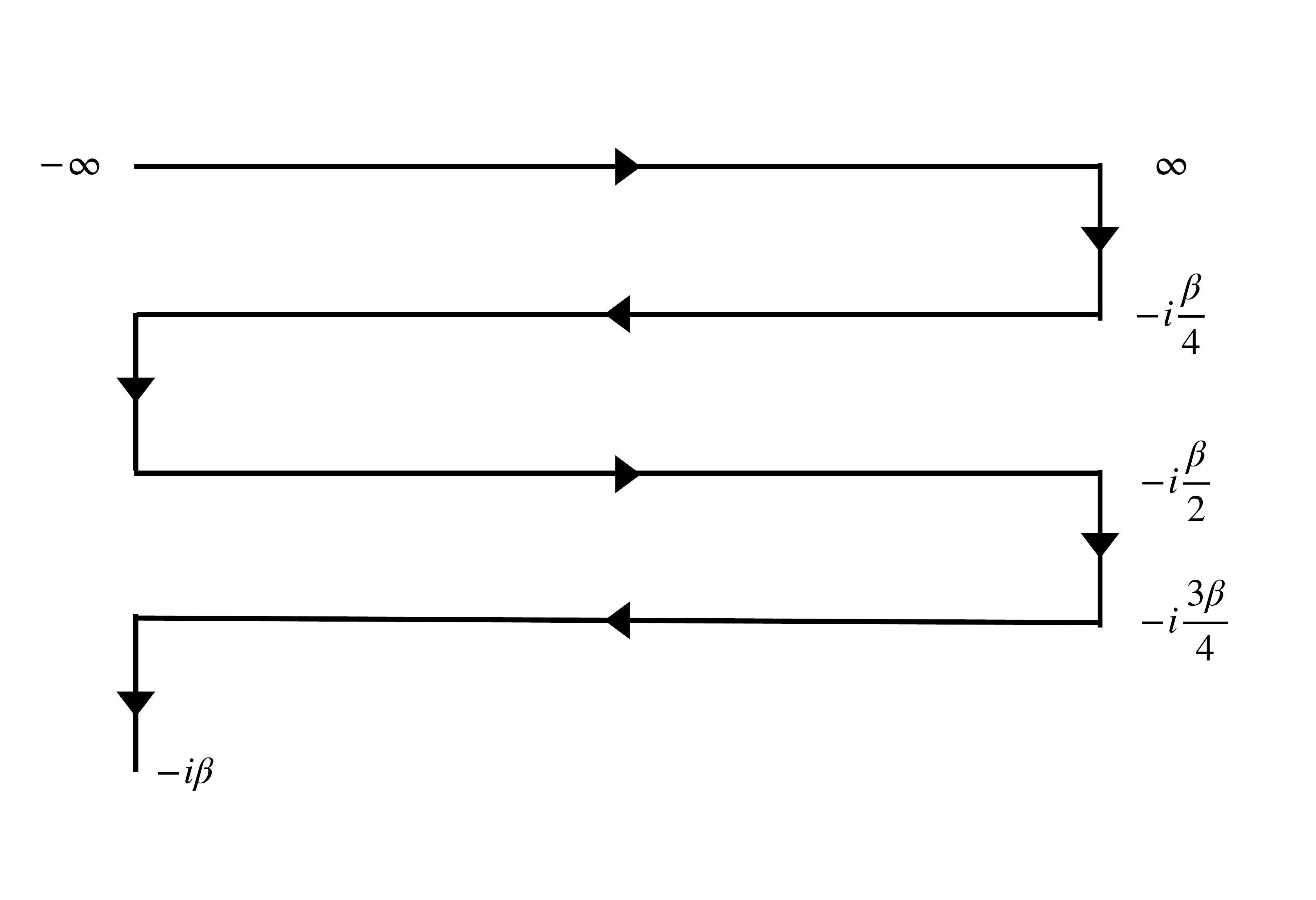}
	\caption{The replicated path integral for OTOC is represented on a SK contour with four real-time branch separated by imaginary time $\beta/4$.}\label{fig:SKContour_S}
\end{figure}

Below $T_d$, while getting cooled in the presence of the bath, the system enters the so-called aging regime where, e.g., the spin-spin correlation $C(t_w,t_w+t)$ depends on the waiting time $t_w$ even for asymptotically large $t_w $. However, by formally taking $t_w\to\infty$ and then coupling to the bath to zero in the saddle-point equations on the SK contour \cite{Cugliandolo2001}, one arrives at solutions identical to the 1-RSB mSG in Sec.\ref{supp-SD-SG-sol}. Hence, to compute OTOC using the SK contour of Fig.\ref{fig:SKContour_S}, rather than taking the above more complicated route of taking $t_w\to \infty$ and then the coupling to bath to zero, one can simply replicate \cite{Houghton1983,Cugliandolo2019} the generating function $\mathcal{Z}$ and take $t_0\to -\infty,~t_f\to\infty$ limits, i.e.
\begin{align}
\mathcal{Z}^n  &= \frac{1}{Z^n}\int \left[\prod_a \mathcal{D}s_a(z)\delta \left(\sum_{i}s^2_{i,a}(z)-N\right)\right]\exp\bigg[\frac{\ci}{\hbar}\int_\mathcal{C} dz \bigg(\sum_{i,a}\frac{M}{2}\left(\frac{\partial s_{i,a}}{\partial z }\right)^2 + \sum_{i_1<...<i_p, a}J^{(p)}_{i_1...i_p}s_{i_1,a}(z)...s_{i_p,a}(z) \bigg) \bigg].
\end{align}
Here $z$ is the variable on the SK contour $\mathcal{C}$ shown in Fig.\ref{fig:SKContour_S}, where the four horizontal real-time branches are denoted as 1,2,3,4 from the top to bottom. We can perform the disorder average over the above replicated generating function. In this case, after taking the $n\to 0$ limit, the resulting time-translation invariant dynamical correlations and responses are identical to that obtained from the 1-RSB saddle point solutions in the mSG phase (Sec.\ref{supp-SD-SG-sol}).
Using the SK contour, the OTOC of Eq.\eqref{eq:OTOCop} can be written as
\beq \label{sup:Faabb}
F_{aabb}(t_1, t_2) = \frac{1}{N^2} \sum_{i j}\langle \overline{ ys^{(4)}_{ia}(t_1)ys^{(3)}_{ja}(0)ys^{(2)}_{ib}(t_2)ys^{(1)}_{jb}(0) }\rangle,
\eeq  
where the average $\langle \dots\rangle$ is with respect to the generating function and the superscripts in $s_{ia}$ refers to the real-time branch. 

\subsection{Kernel equation for OTOC in the PM and marginal SG phase}\label{supp-SG-otoc}

We first discuss the real-time retarded and Wightmann functions, which we need to obtain OTOC using the  Schwinger-Keldysh contour in Fig.\ref{fig:SKContour_S}. 
The retarded function is obtained via analytical continuation as 
\beq 
Q^R_{ab}(\w)=Q_{ab}(\ci\w_k\rightarrow \w+\ci 0^+)=\qr^R(\w)\delta_{ab}.
\eeq 
We note that the $\qe\epsilon_{ab}$ term in $Q_{ab}(\w_k)$ [Eq.\eqref{rsb1}] does not contribute to the retarded function, which is replica diagonal. The spectral representation of $Q_{ab}(\tau)$ is given by the relation Eqs.\eqref{sp:qrspectral} and \eqref{sp:Qabtau}, namely
\beq \label{sp:Qabtauspect}
Q_{ab}(\tau) = \qr(\tau)\delta_{ab}+\qe\epsilon_{ab}=\qe\epsilon_{ab}-\delta_{ab}\int_{-\infty}^\infty d\w \rho(\w)n_B(\w)e^{\w (\beta-\tau)}
\eeq 
The Wightmann function is defined as $Q^W_{ab}(t):=Q_{ab}(\tau=\ci t + \beta/2)$. Now, using the above Eq.\eqref{sp:Qabtauspect}, we obtain the Wightmann function after doing Fourier transformation as follows
\beq \label{sup:Qw-aa}
Q^W_{ab}(\w) = \bigg[ \epsilon_{ab} 2\pi \delta(\w)\qe +  \delta_{ab}\qr^W(\w)\bigg], \hspace{0.5cm}\qr^W(\w)=-\frac{\pi \rho(\w)}{\sinh(\beta\w/2)}
\eeq 
The replica diagonal component of Wightmann function contains Edward-Anderson parameter $\qe$ as well as the regular component $\qr^W$. The replica off-diagonal component of $Q^W$ contains only Edward-Anderson term $\epsilon_{ab} 2\pi \delta(\w)\qe$. 

The exponential growth $\sim e^{\lambda_\mathrm{L}t}$ appears in the OTOC $F_{aabb}(t_1,t_2)$ [Eq.\eqref{sup:Faabb}] at $\mathcal{O}(1/N)$, namely $F_{aabb}(t_1,t_2)=\mathcal{O}(1)+(1/N)\mathcal{F}_{aabb}(t_1,t_2)$ with $\mathcal{F}_{aabb}(t,t)\sim e^{\lambda_\mathrm{L}t}$. The quantity $\mathcal{F}_{aabb}(t_1,t_2)$ can be obtained via the ladder series \cite{KitaevKITP,Maldacena2016,BanerjeeAltman2016} shown in Fig.\ref{fig:LadderDiagram}(b) of the main text, i.e. the kernel equation
\begin{align}
\mathcal{F}_{aabb}(t_1,t_2)=\int dt_3 dt_4\sum_{c} K_{aacc}(t_1,t_2,t_3,t_4)\mathcal{F}_{ccbb}(t_3,t_4)
\end{align}
In the intermediate but long time scale of chaotic growth regime $\lambda_\mathrm{L}^{-1}\lesssim t\lesssim \lambda_\mathrm{L}^{-1}\ln(N)$, the propagators along the horizontal lines from $t_1$ to $t_3$ and from $t_2$ to $t_4$ in Fig.\ref{fig:LadderDiagram}(b) (main text) can be approximated by the retarded propagator \cite{KitaevKITP}. In this case, since the retarded propagator is replica diagonal both in the PM and SG phases, the above equation reduces to 
\beq \label{sup:Fa}
\mathcal{F}_{a}(t_1, t_2) = \int dt_3 dt_4 K_a(t_1, t_2, t_3, t_4)\mathcal{F}_a(t_3, t_4)
\eeq  
with $\mathcal{F}_a=\mathcal{F}_{aaaa}$. Here the kernel for $p=3$ is $K_a(t_1,t_2,t_3,t_4)=3J^2Q^R_{aa}(t_{13})Q^R_{aa}(t_{24})Q^W_a(t_{34}) $, where $t_{ij}=t_i-t_j, i,j=1, 2, 3, 4$ and $Q^R(t_{ij})$ is retarded propagator and $Q^W(t_{34})$ is Wightmann function in real time. In the chaos regime using exponential growth ansatz, $\mathcal{F}_a(t_1,t_2)=f_a(t_1,t_2)e^{\lambda_L(t_1+t_2)/2}$, and doing Fourier transformation of Eq.\eqref{sup:Fa}, we obtain 
\beq \label{sup:fa}
f_a(\w) = \int \frac{d\w'}{2\pi} K_a(\w, \w')f_a(\w')
\eeq 
with the kernel in the frequency domain 
\beq \label{sup:Ka-w}
K_a (\w, \w') = 3J^2 Q^R_{aa}\left(\w+\ci\frac{\lambda_L}{2}\right)Q^R_{aa}\left(-\w+\ci\frac{\lambda_L}{2}\right)Q^W_{aa}(\w-\w').
\eeq 
We can think of the Eq.\eqref{sup:fa} as an eigenvalue equation in the frequency space. We discuss in next section the numerical diagonalization of the kernel to obtain Lyapunov exponent.  


The kernel in PM phase is different from SG phase as the Wightmann function in PM phase doesn't contain Edward-Anderson term $\epsilon_{ab} 2\pi \delta(\w)\qe$ since $\qe = 0$ in PM phase. 
Therefore, the information of the PM phase and SG phase is explicitly encoded in the Wightmann function, and, of course, implicitly in the retarded function, appearing in the kernel.

\subsection{Numerical diagonalization of kernel}\label{supp-kernel-diag}
We can view Eq.\eqref{sup:fa}) as an eigenvalue equation,  $\int \frac{d\w'}{2\pi} K_{a}(\w,\w') f_a(\w') = \lambda f_a(\w)$, with eigenvalue $\lambda=1$. 
The kernel Eq.\eqref{sup:Ka-w} is not symmetric with respect to frequencies ($\w$, $\w'$). We symmetrize the kernel using  particle-hole symmetry of the Hamiltonian. Using this symmetry it is easy to show that 
\begin{equation}
Q^R_{aa}\left(\omega+\frac{\ci\lambda_L}{2}\right)Q^R_{aa}\left(-\omega+\frac{\ci\lambda_L}{2}\right)= \left\vert Q^R_{aa}\left(\omega+\frac{\ci\lambda_L}{2}\right)\right\vert^2
\end{equation} 		          
We can now define a symmetric kernel by defining  $f_a(\omega)=\vert Q^R_{aa}(\omega + \ci \lambda_L/2)\vert\tilde{f}_a(\omega)$ and obtain the symmetric kernel as
\begin{equation} \label{eq: kernel-sym}
\tilde{K_a}(\w, \w') = 3J^2\left\vert Q^R_{aa}\left(\w + \frac{\ci \lambda_L}{2}\right)\right\vert  {Q^W_{aa}(\w-\w')} \left\vert Q^R_{aa}\left(\w' +  \frac{\ci \lambda_L}{2}\right)\right\vert
\end{equation}		   
Hence the eigenvalue equation becomes 
\beq 
\int_{-\infty}^{+\infty} \frac{d\w'}{2\pi} \tilde{K_a}(\w, \w') \tilde{f}_a(\w') = \lambda \tilde{f}_a(\w)
\eeq 
To solve the above eigenvalue equation we first discretize frequency, namely $\w_n$, $n$ being integer, ranging from $-\w_{\text{max}}$ to $\w_{\text{max}}$ with spacing $2\w_{\text{max}}/N_\w$, where $N_\w$ is total number of discretized frequency points and $\w_\mathrm{max}$ is a upper frequency cutoff. We diagonalize the matrix $\tilde{K}_a$ and find the value of $\lambda_\mr{L}$ such that $\tilde{K}_a$ has at least has one eigenvalue $\lambda=1$. 

\section{Spectral function in the cPM, qPM and mSG phases and across the crossover from strong to weak chaos}\label{supp-spect}
In this section we discuss the evolution spectral function as a function of $T$ and $\hbar$ over the phase diagram in Fig.\ref{fig:PhaseDiagram}. Thermodynamically the system can be characterized as gapped or gapless only at low temperature ($T\ll J,\Gamma$) based on the temperature dependence of specific heat or susceptibility. As discussed in Ref.\onlinecite{Cugliandolo2001}, the qPM phase is strongly gapped with specific heat $C_v\sim e^{-E_g/T}$ and gap $E_g$, whereas the mSG phase is gpaless with $C_v\sim T$ at low temperature. Here, however, we are also interested to study chaos at intermediate temperatures. Hence to relate the nature of the excitations in the system with chaos we look into the spectral function $\rho(\omega)$ defined in Sections \ref{supp-SD-PM} and \ref{supp-SD-SG-sol} for the PM and SG phases, respectively.

\begin{figure}[htb]
	\centering
	\includegraphics[width=0.5\textwidth]{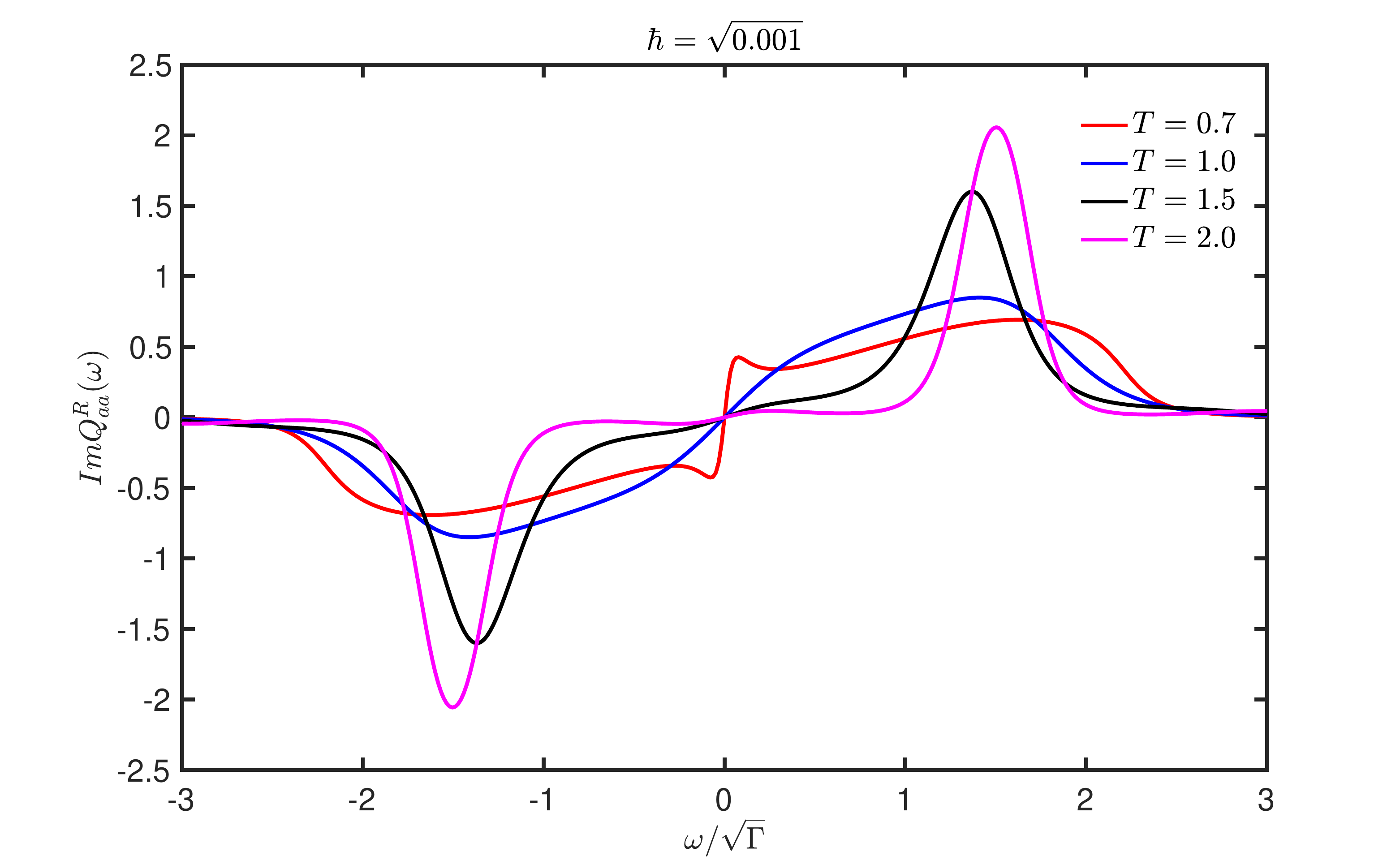}
	\caption{The imaginary part of $Q^R_{aa}(\w)$ [$\text{Im}Q^R_{aa}(\w)=-\pi \rho(\w)$]  as a function of $\w/\sqrt{\Gamma}$ is shown for different $T$ at $\hbar=\sqrt{0.001}$ in the cPM phsae. The spectral function is gapless for $T\gtrsim 0.7$ and becomes progressively more (soft) gapped at higher temperatures. }
\end{figure}
\begin{figure}[htb]
	\centering
	\includegraphics[width=0.5\textwidth]{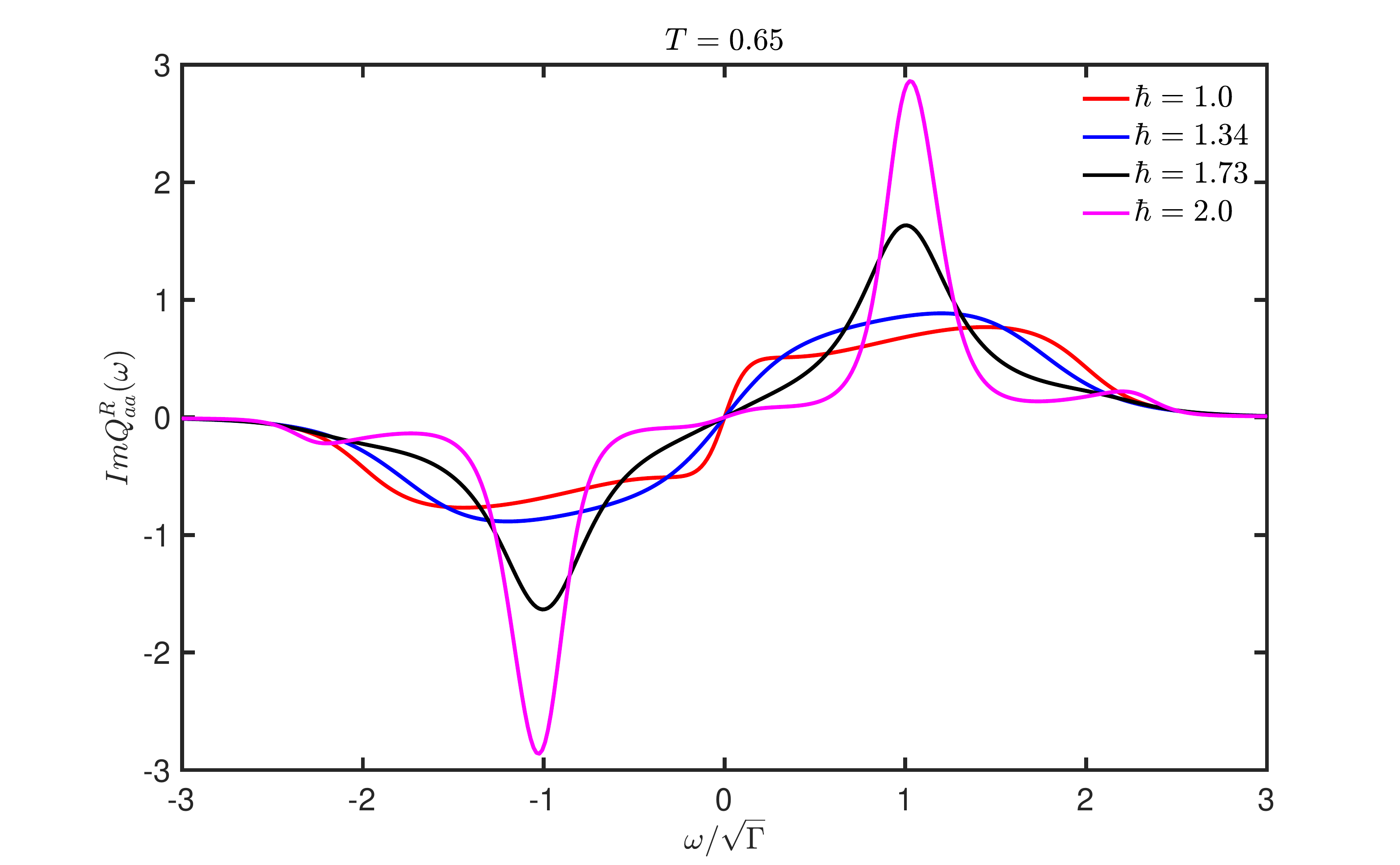}
	\caption{The imaginary part of $Q^R_{aa}(\w)$ [$\text{Im}Q^R_{aa}(\w)=-\pi \rho(\w)$] as a function of $\w/\sqrt{\Gamma}$ is shown for different quantum fluctuation $\hbar$ at temperature $T=0.65$ close to $T_d$ in the cPM phase. The spectral function becomes more gapped with increasing $\hbar$.}
\end{figure}
\begin{figure}[htb]
	\centering
	\includegraphics[width=0.5\textwidth]{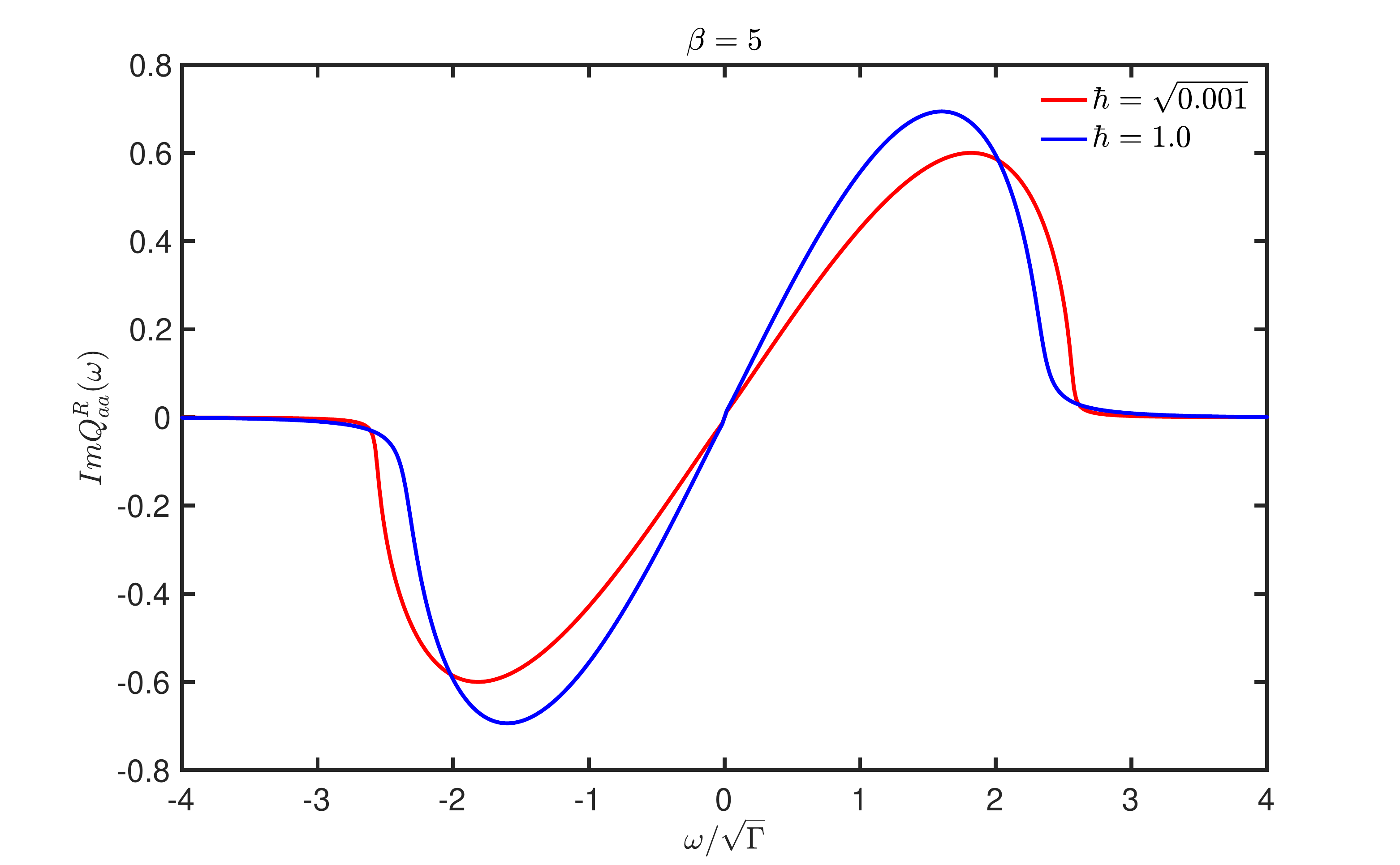}
	\caption{$\text{Im}Q^R_{aa}(\w)=-\pi \rho(\w)$ vs $\w/\sqrt{\Gamma}$ is shown for different quantum fluctuation $\hbar$ at inverse temperature $\beta=5$ in the mSG phase. The spectral function remains gapless throughout the mSG phase.}
\end{figure}
 
For bosonic systems $\omega \text{Im}Q^R(\omega)\geq 0$ and thus the spectral function either approaches zero or has a discontinuity at $\w=0$. We operationally characterize, albeit in somewhat \emph{ad hoc} way, whether the system is gapped (or soft gapped) or gapless as follows. The gapped free retarded propagator in this model for $J_p=0$ is $Q^R_0(\w) =-\G/[({\w}+\ci\eta)^2-(E^0_g)^2]
$
where $E^0_g = \sqrt{\G z}$ is the gap, where $z$ is Lagrange multiplier that appears in Eq.\eqref{eq:SaddlePoint} and $\eta>0$ is a small broadening.  
Hence, the free spectral function $\rho^0(\omega)=-(1/\pi)\text{Im} Q^R_0(\w)$ can be written as 
\beq\label{sup-eq-free-spect}
\rho^0(\w) = \frac{\G}{2\pi E^0_g}\bigg[ \frac{\eta}{({\w}+E^0_g)^2+\eta^2} -\frac{\eta}{({\w}-E^0_g)^2+\eta^2}\bigg] 
\eeq
Therefore, we see that the gapped free spectral function has peaks at $\w=\pm E^0_g$ with Lorentzian decay around these peaks. We can estimate the spectral weight at $\w\rightarrow 0$ as 
$
\rho_{0}=\rho^0(\w\to 0) =(2\eta\w\G)/(\pi(E^0_g)^4)$. Numerically we can take $\eta$ as the spacing of discrete frequency points i.e $\Delta\w$. 
The slope of spectral function at $\w\rightarrow 0$ is therefore
$
A_0= (\partial\rho^0(\w)/\partial\w)\vert_{\w=0} =2\eta\G/[\pi (E^0_g)^4]$.

\begin{figure}[htb]
	\centering
	\includegraphics[width=0.6\textwidth]{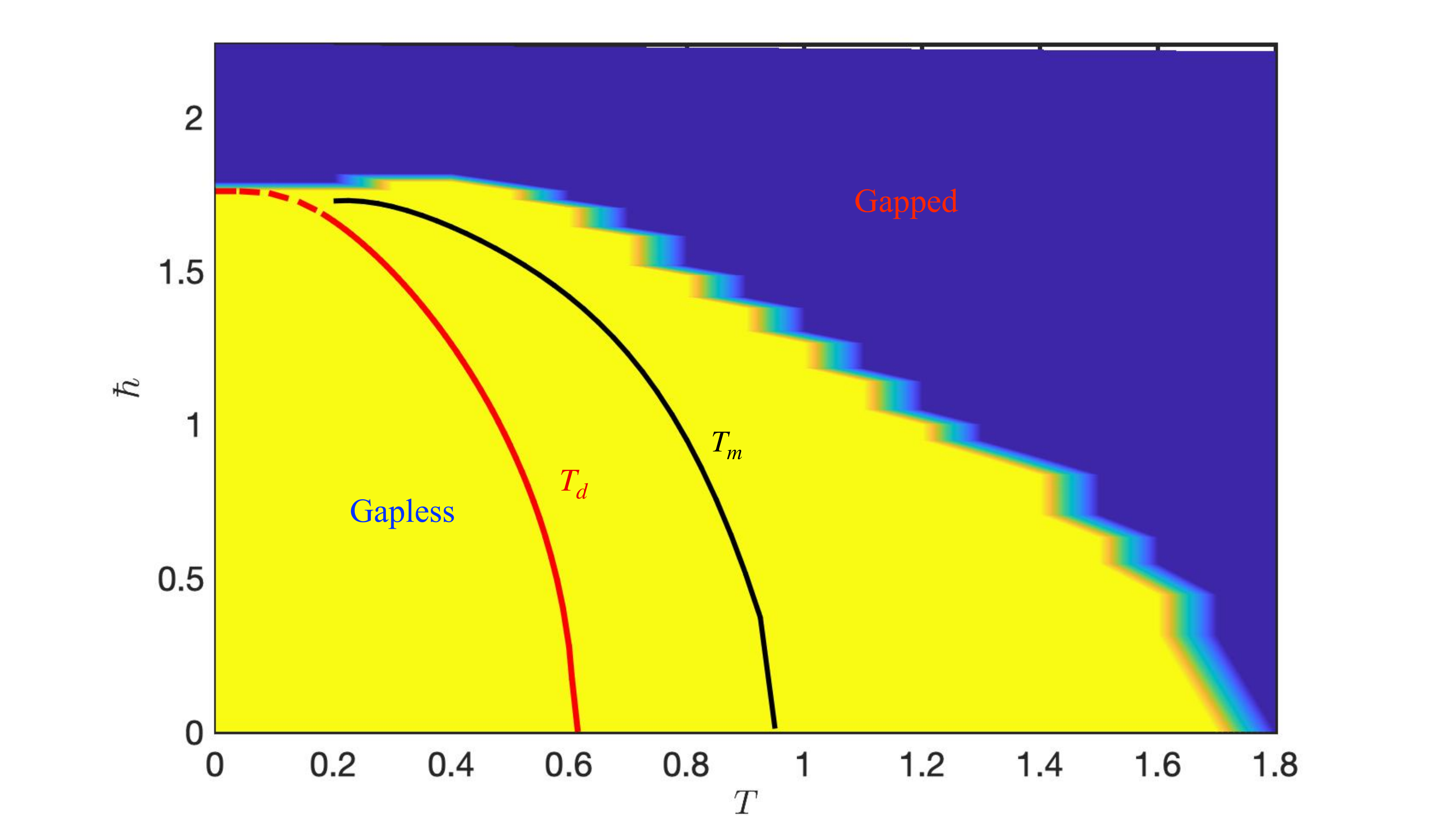}
	\caption{The phase diagram based on nature of spectral function, i.e. gapped (blue region) or gapless (yellow region), in the $p=3$ spin glass model. $T_d$ denotes dynamical transition line between mSG and PM phase. $T_m$ is temperature where $\lambda_\mathrm{L}$ has a broad minimum. 
	}
	\label{fig:sup-fig-5}
\end{figure}

After numerically calculating the spectral function from the solution of the saddle point equations, we compute the slope of $\rho(\w)$ at all discrete frequency points between the peak at $\w=-E_g$, where $\rho(\w)$ has maximum value, and $\w=0$. Due to particle-hole symmetry of the model, $\rho(\w)$ has odd parity and hence we do not need to check the slopes for $\w>0$. We count the number of points where $(\delta \rho(\w)/\delta \w)\vert_{\w=0}<A_0$. 
If the number of counts is greater than $10$, then we define the spectral function as gapped otherwise it is characterized as gapless.
Using this procedure we mark the gapless and gapped regions in the phase diagram in Fig: \ref{fig:sup-fig-5}. 
The spectral function is always gapless in the mSG phase. The systems is gapped in the qPM phase and, for finite $\Gamma$, the system becomes gapped at high temperatures in the cPM phase. The spectrum is expected to be gapless strictly at the classical limit $\Gamma=0$ \cite{Cugliandolo2001}. We note that the nature of the spectrum does not undergo any qualitative change across the crossover from strong to weak chaos around $T_m(\hbar)$.

The nature of the excitation spectrum can be understood easily at large $\Gamma$ or $T$ where the system effectively becomes non-interacting. For the $\eta\rightarrow 0$ limit the free spectral function $\rho^0(\w)$ in Eq.\eqref{sup-eq-free-spect} has two delta function peaks, i.e. $\rho^0(\w)=\sum_{s=\pm} (s\Gamma/2E_g^0)\delta(\w+sE^0_g)$. Using the sum rule condition of Eq.\eqref{eq:PM-18}, we obtain 
$\tanh(\beta E^0_g/2)=\Gamma/2E^0_g$. At very low temperatures ($\beta\rightarrow \infty$), $\tanh(\beta E^0_g/2)\simeq 1$ and thus $E^0_g=\Gamma/2$. Due to the large gap, the perturbative effect of self-energy is very small for $\Gamma\gg T,J$ i.e. in the qPM phase. On the other hand, in the high temperature limit ($\beta\rightarrow 0$) $\tanh(\beta E^0_g/2)\simeq \beta E^0_g/2$ and hence $E^0_g=\sqrt{\Gamma T}\ll T$ for $T\gg \Gamma$. The self-energy effect will lead to the broadening of the peak, however the gap $E_g\approx E_g^0$ for $T\gg J$.

\section{Lyapunov exponent}\label{supp-lyapunov}
\subsection{$\lambda_\mathrm{L}(T)$ in the cPM phase at high temperature}\label{supp-highT-fit}
\begin{figure}[htb]
\centering
	\includegraphics[width=0.5\textwidth]{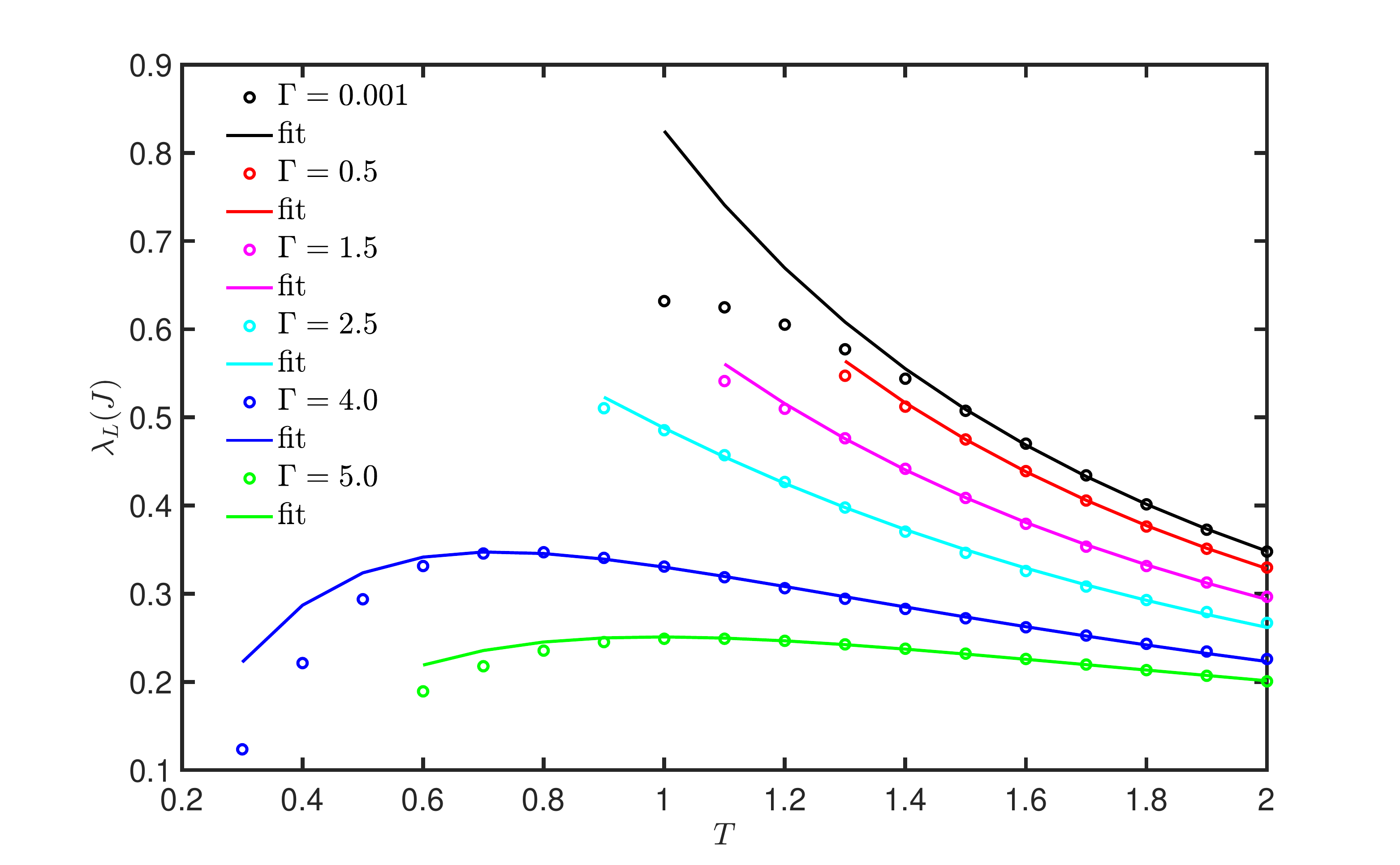}
	\caption{The fitting of $\lambda_L$ (in units of $J$) with  $\frac{A}{T^{2}}\exp(-b/\sqrt{T})$. 
	}\label{sup-fig-7}
\end{figure}

As shown in Fig.\ref{fig:lambdahbar}(a) (inset) in the main text, $\lambda_\mathrm{L}(T)$ decreases with temperature for $T>T_m$. 
As discussed in the preceding section, at high temperature ($T\gtrsim J$) in the cPM phase, the system has a spectral gap $E_g\approx \sqrt{\Gamma T}$, which is not affected much by interaction in this temperature regime. Motivated by this, we fit the temperature dependence of $\lambda_\mathrm{L}$ with $(A/T^a)\exp{(-b/\sqrt{T})}$ for various $\Gamma$ values ranging from the classical ($\Gamma \to 0$) to the quantum limit $\Gamma>T$, as shown in Fig.\ref{sup-fig-7}(a). We find $a\simeq 2$ implying that $\lambda_\mr{L}$ decreases as $1/T^2$ at high temperature. 

\subsection{$\lambda_\mathrm{L}(T)$ in the PM phase at low temperature}\label{supp-lowT-fit}
\begin{figure}[htb]
    \centering
    \begin{tabular}{cc}
	\includegraphics[width=0.5\textwidth]{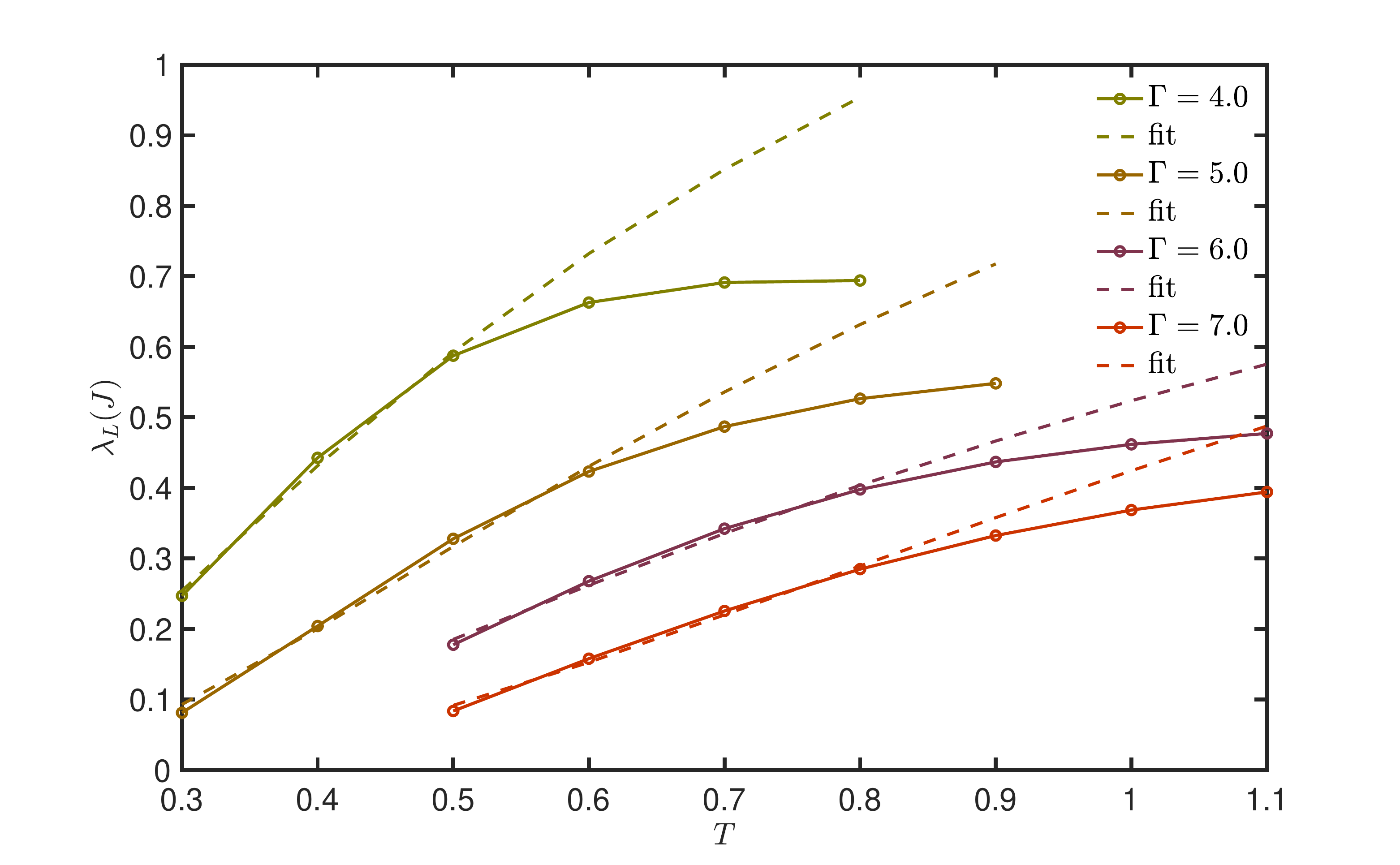} &
	\includegraphics[width=0.5\textwidth]{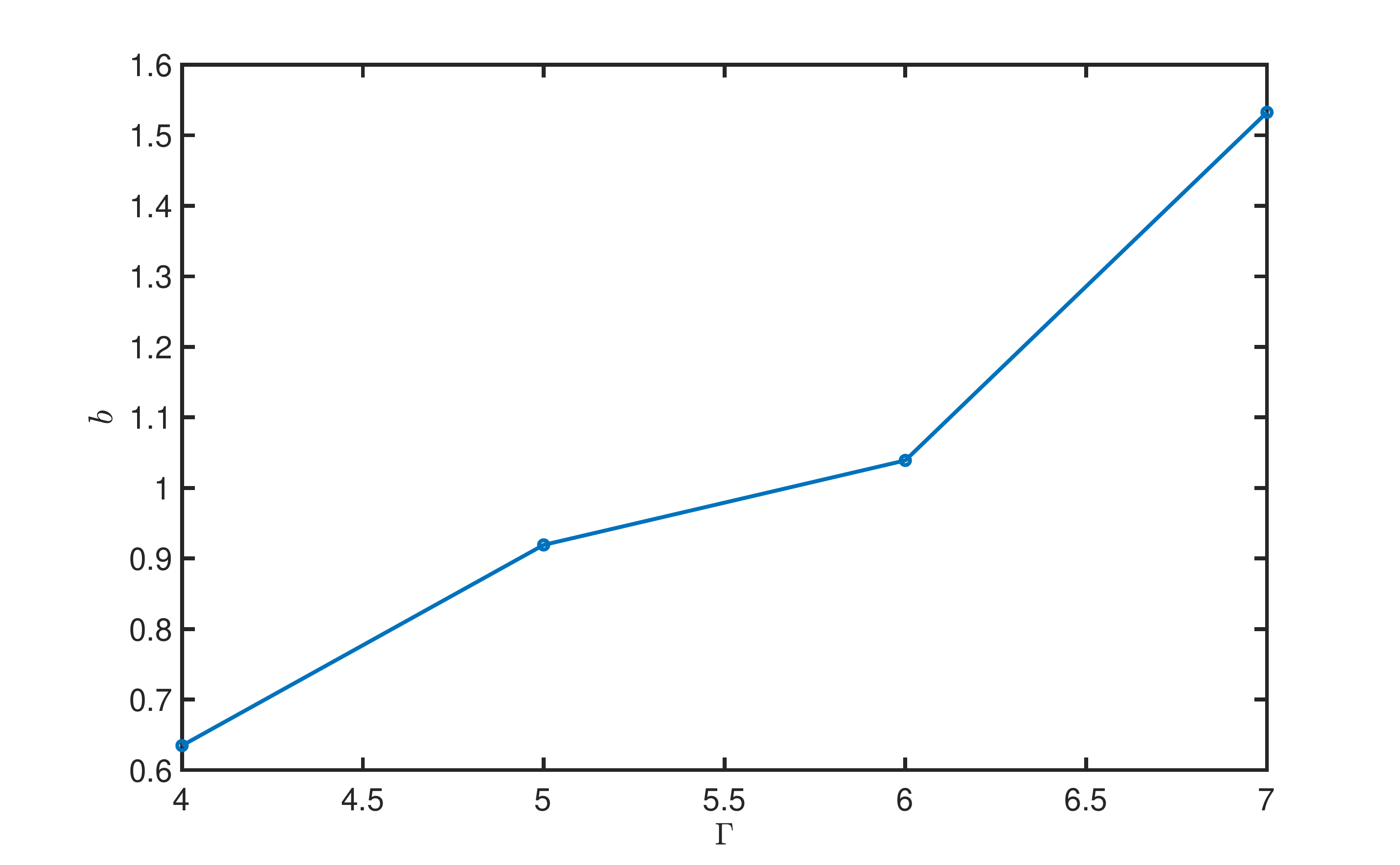}
	\end{tabular}
		\caption{The fitting of $\lambda_L$ (in units of $J$) with $\lambda_L={A}\exp(-b/T)$ for $\Gamma \gg T,J$ in the left panel for sevral values of $\Gamma$. The right panel shows the fitting parameter $b$ as a function of ${\Gamma}$. We find $b \propto {\Gamma}$ as expected from the non-interacting limit. }
	\label{sup-fig-8}
\end{figure}

As discussed earlier, the spectral function becomes strongly gapped for $\Gamma\gg T,J$ with a very weak self energy effect.
Therefore, we expect the Lyapunov exponent to decrease exponentially as $\sim \exp(-E_g/T)$ with $E_g \simeq E_g^0=\Gamma/2$. 
We fit $\lambda_L$ with $ A \exp(-b/T)$, where $A$ and $b$ are fitting parameters. The results for the fitting and  the extracted values of the parameter $b$ are shown in Fig.\ref{sup-fig-8} for several values of $\Gamma\gg T$. 

\subsection{$\lambda_L(T)$ in the mSG phase}\label{supp-lambdaL}
\begin{figure}[htb]
	\centering
	\includegraphics[width=0.5\textwidth]{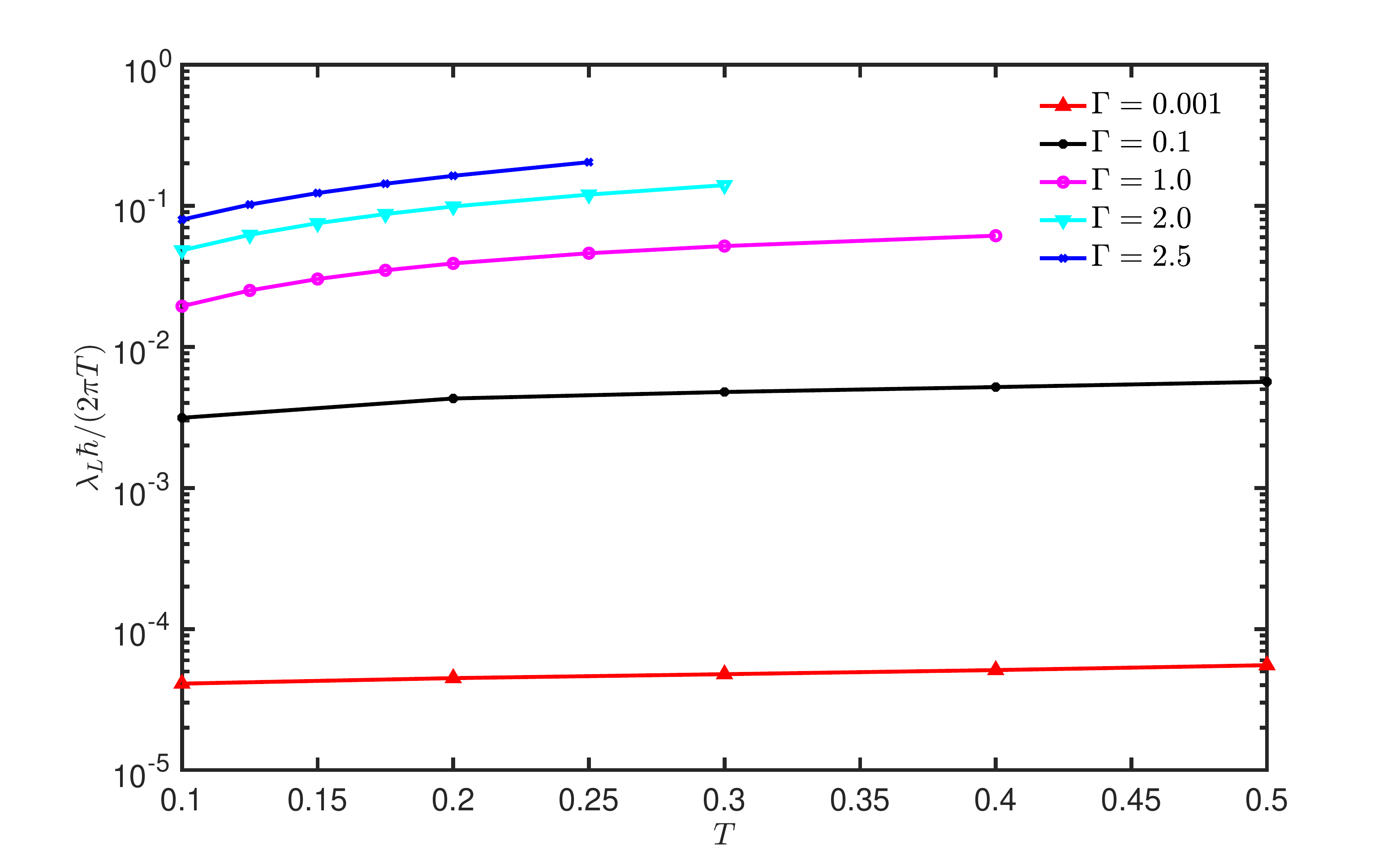}
	\caption{$\lambda_L\hbar/2\pi T$ vs $T$ is shown for different $\Gamma$ in the mSG phase.}\label{sup-fig-9}
\end{figure}

As discussed in the main text, the Lyapunov exponent follows a power-law temperature dependence $\lambda_\mathrm{L}\sim T^\alpha$ in the mSG phase at low temperature with $\alpha$ varying between $2$ to $1$. In particular, in the classical limit $\hbar \to 0$, $\lambda_\mathrm{L}$ has a linear temperature dependence, analogous to that expected from the chaos bound $\lambda_\mathrm{L}^{(b)}=2\pi T/\hbar$. For the latter, the linear-$T$ coefficient diverges as $1/\hbar$ in the classical limit. On the contrary, the linear-$T$ coefficient of $\lambda_\mathrm{L}$ in the mSG phase approaches a finite constant in the classical limit, implying $\lambda_\mathrm{L}\hbar/2\pi T \to 0$ for $\Gamma\to 0$, as shown in Fig.\ref{sup-fig-9}. We also plot $\lambda_\mathrm{L}/\lambda_\mr{L}^{(b)}$ for a few other values of $\Gamma$ for comparison.

\subsection{Perturbative analysis in the mSG phase}\label{supp-perturbative-lamb}
We have shown in Sections \ref{supp-SD-SG-sol} and \ref{supp-SG-otoc} that the saddle point equations and the kernel equation for OTOC in the mSG phase have forms similar to those in the PM phase of the model [Eq.\eqref{eq:Hamiltonian}] with $p=2$ and $p=3$ terms with couplings $J_2=\sqrt{3\qe} J$ and $J_3=J$. Here, we obtain $\lambda_\mathrm{L}(T)$ by treating the $p=3$ term as perturbation around the $p=2$ saddle point in the low-temperature quantum limit $T\ll \Gamma$. The analysis is similar to that done in the Fermi liquid (FL) phase \cite{Kim2020,Kim2020a} of a large-$N$ fermionic model related to the SYK model. The perturbative analysis is well controlled, even though the ratio $J_3/J_2 = \sqrt{3\qe} $ is not necessarily small, since the $p=3$ spin interaction term  is irrelevant at low energies. 

The kernel equation for the PM phase of the $p=2+p=3$ model can be written as 
\begin{align} \label{supp-K}
K &= K_1 K_2 \\
K_2 f(\w) &= \left\vert Q^R\left(\w+\frac{\ci \lambda_L}{2}\right)\right\vert^2 f(\w) \\
K_1 f(t) &= [J_2^2 + 3J_3^2 Q^{W}(t)]f(t)
\end{align}
where $Q^R$ and $Q^W$ are retarded and Wightmann functions, respectively. The kernel $K_1$ is diagonal in the time domain and $K_2$ is diagonal in the frequency domain. 

To calculate $\lambda_L$ we need to obtain $|Q^R|^2$ and $Q^W$. The saddle point equation in real frequency is given by 
\beq \label{supp:SD-p2p3}
(Q^R)^{-1}(\w) = -\frac{\w^2}{\G}+z-J_2^2 Q^R(\w)-\Sigma^R_3(\w) \\
\eeq 
where $\Sigma^R_3(\w)$ is the self-energy in real frequency for $p=3$ spin interaction term.  The self-energy in imaginary time is given as $\Sigma_3(\tau) =(3/2) J_3^2[Q(\tau)]^2$. When $J_3=0$, the saddle-point equation above can be solved exactly. The spectral function can be made gapless, like in the mSG phase, by choosing $z=2J_2$. In this case the solution is given by
\beq 
Q^R_0(\w) = \frac{2}{J_2}\bigg(\frac{1}{2}-\tilde{\w}^2+\ci \tilde{\w}\sqrt{1-\tilde{\w}^2} \bigg)
\eeq 
where $\tilde{\w}=\frac{\w}{2\sqrt{J_2\G}}$.
At low energies $\tilde{\w}<<1$, using $\sqrt{1-\tilde{\w}^2} \simeq 1$ we get 
\beq 
Q^R_0(\w) =\frac{2}{J_2}\bigg(\frac{1}{2}-\tilde{\w}^2+\ci \tilde{\w} \bigg) = q_1^0 + \ci q_2^0,
\eeq 
where $q^0_1$ and $q^0_2$ are real and imaginary parts of $Q^R_0(\w)$, respectively, leading to
\beq 
\rho^0(\w) = -\frac{1}{\pi}\text{Im}Q^R_0(\w)=-\frac{\w}{\pi J_2\sqrt{J_2\G}}
\eeq  
To calculate the Lyapunov exponent perturbatively, we need to compute $|Q_R|^2$ and $Q_W$ perturbatively around the $J_3=0$ gapless solution. 
For $|Q^R|^2$, the zeroth order $J_3=0$ solution is not enough and we have to incorporate the quasi-particle decay contribution, coming from the imaginary part of self-energy term  $\Sigma_3(\tau)$. The imaginary part of $\Sigma^R_3(\w)$ \cite{coleman2015} can be computed perturbatively using the zeroth order solution. At the leading order and small frequency we get
\beq 
\text{Im}\Sigma^R_3(\w) \simeq \frac{J_3^2}{2\pi J_2^3\G}\bigg[ \frac{\w^3}{2}+2\pi^2 \w T^2\bigg] + \mathcal{O}(\w^5)
\eeq 
Using the saddle-point equation [Eq.\eqref{supp:SD-p2p3}] and $\Sigma^R_3(\w)$ above, we obtain 
\beq \label{supp:QR}
J_2^2 \left\vert Q^R\left(\w + \ci \frac{\lambda_L}{2}\right)\right\vert^2 = 1+\frac{1}{J_2^2 q^0_2}\left[ -\frac{\w \lambda_L }{\G} - \frac{J_3^2}{2\pi J_2^3 \G}\left(\frac{\w^3}{2}+2\pi^2 T^2 \w \right) \right]
\eeq 
 For $Q^W$, the $J_3=0$ solution is sufficient and the Wightmann function $Q^W(t)=Q(it+\beta/2)$ in real frequency is given by
\beq \label{supp:Qw-t}
Q^{W}(\w) = - \frac{\pi \rho^0(\w)}{\sinh[\beta\w/2]} =  \frac{\w}{J_2\sqrt{J_2\G}{\sinh[\beta\w/2]}}
\eeq 
By doing Fourier transformation we obtain 
\beq 
Q^{W}(t) = \frac{\pi T^2}{J_2\sqrt{J_2\G}}\sech^2(T\pi t)
\eeq 
Using equations \eqref{supp:QR} and \eqref{supp:Qw-t}, the kernel equation \eqref{supp-K} is written as
\beq 
K= 1-\frac{\lambda_L}{\sqrt{J_2\G}} - \frac{J_3^2 \pi T^2}{J_2^3\sqrt{J_2\G}} + \frac{3J_3^2}{2\pi J_2^3 \sqrt{J_2\G}}\partial_t^2 + \frac{3J_3^2 \pi T^2}{J_2^3 \sqrt{J_2\G}}\sech^2(t\pi T)
\eeq 
Using rescaled variable $s=t\pi T$ and equating the maximum eigenvalue of $K$ to 1, we obtain a simplified expression 
\beq 
\bigg(-\frac{1}{2}\frac{\partial^2 }{\partial s^2} - \sech^2 s \bigg) f(s) = -\frac{1}{3}\bigg( \lambda_L \frac{J_2^3}{\pi J_3^2 T^2} + 1\bigg) f(s)
\eeq 
The bracketed term in the left-hand side of the above equation is the Schr\"{o}dinger Hamiltonian with P\"{o}schl-Teller potential \cite{Poschl1933} whose eigenvalues are well known. The ground-state energy eigenvalue which maximizes $\lambda_L$ is $-1/2$. Using this, we obtain the Lyapunov exponent 
\beq \label{sup-eq-67}
\lambda_L \simeq \frac{\pi J_3^2 T^2}{2 J_2^3}
\eeq 

\begin{figure}[htb]
	\centering
	\includegraphics[width=0.5\textwidth]{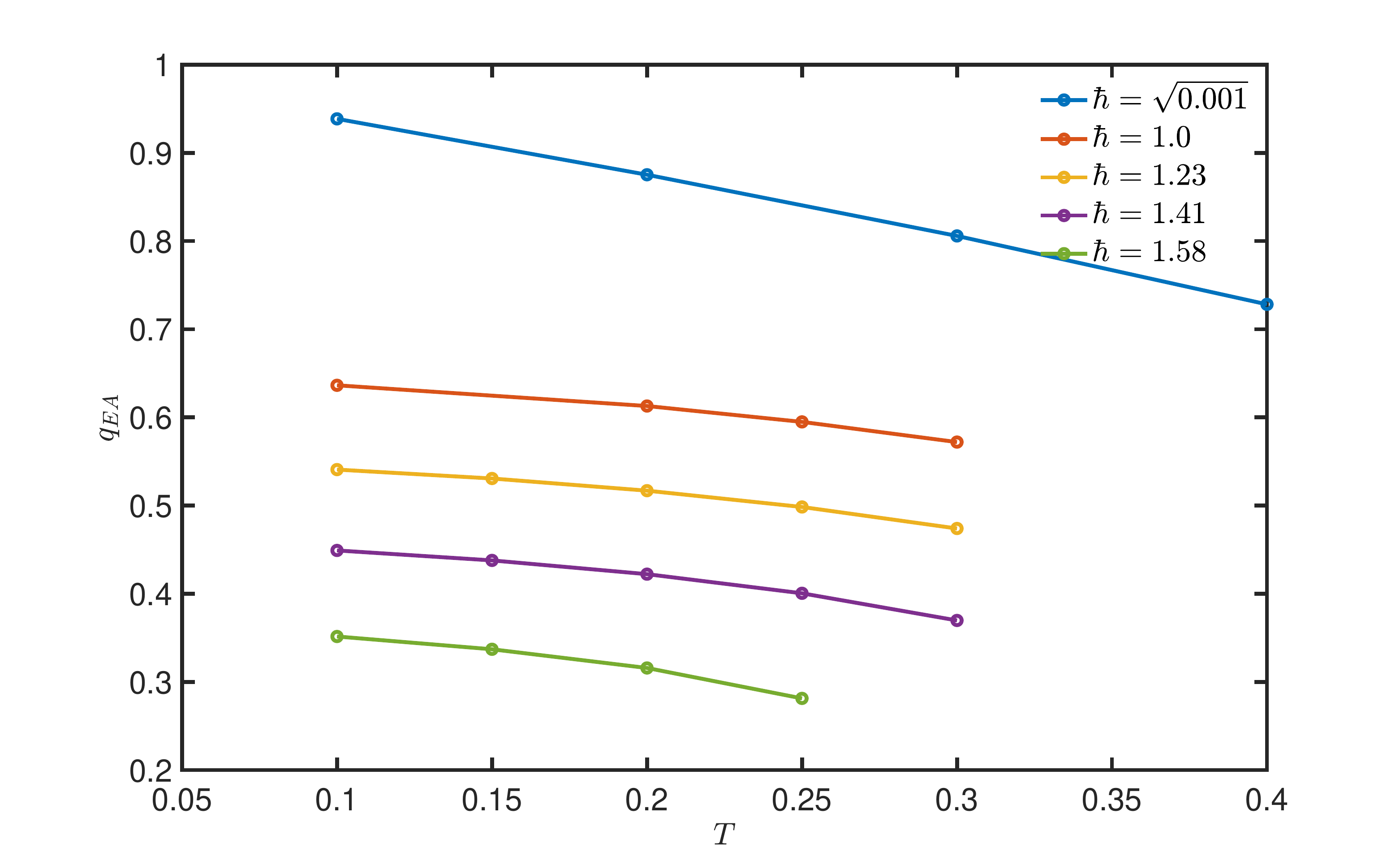}
	\caption{The Edward-Anderson parameter $\qe$ is shown as a function of temperature ($T$) for several values of quantum fluctuation $\hbar$ ranging from the classical to quantum limit.}
	\label{sup-fig-10}
\end{figure}

From Fig:\ref{sup-fig-10} we see that the coefficient $J_2 \propto \sqrt{\qe}$ has weak dependence on temperature for low temperature regime in the quantum limit $\hbar \gtrsim  1$. Therefore, due to quasi-particle decay as in a Fermi liquid\cite{BanerjeeAltman2016,Kim2020,Kim2020a}, we also find the Lyapunov exponent $\lambda_\mathrm{L} \propto T^2$ in the quantum limit at low temperature. This analytical result agrees well with the numerically obtained power-law form $\lambda_\mathrm{L}\sim T^{\alpha}$ in Fig.\ref{fig:lambdahbar}(d) in the main text. From numerical fitting, we find that the exponent $\alpha$ is between $1.8$ to $2$ for $\hbar\gtrsim 1$.  

On the contrary, we find $\lambda_L\propto T^{1.1} \sim T $ [Fig\ref{fig:lambdahbar}(d), main text] in the classical limit $(\hbar\ll 1)$ for the low temperature regime. The above analytical calculation therefore does not hold in the classical regime. There are a few possible reasons behind this. Our perturbative analysis in the mSG phase is done by drawing analogy to the $p=2+p=3$ PM phase at the level of the saddle point and the OTOC kernel. However, the sum rule condition is $-\int d\w \rho(\w)n_B(\w)= 1$ for the PM phase in contrast to the mSG phase where $-\int d\w \rho(\w)n_B(\w)= 1-\qe$
. The violation of this sum rule is small for $\hbar\gtrsim 1$ since $\qe<1$, as seen can be seen in Fig:\ref{sup-fig-10}, and the analytical result agrees with the numerical result for $\alpha$ here. However, for $\hbar\lesssim 1$, the sum rule violation is large and the temperature dependence of $\qe$ is also stronger. Presumably due to these, the analytical result ($\lambda_\mathrm{L}\propto T^2$) in this section is not applicable for the classical limit where $\lambda_L \propto T$.

\section{Correlation function and the onset of two-step relaxation }\label{supp-twostep-corr}
\begin{figure}[htb]
	\centering
	\includegraphics[scale=0.5]{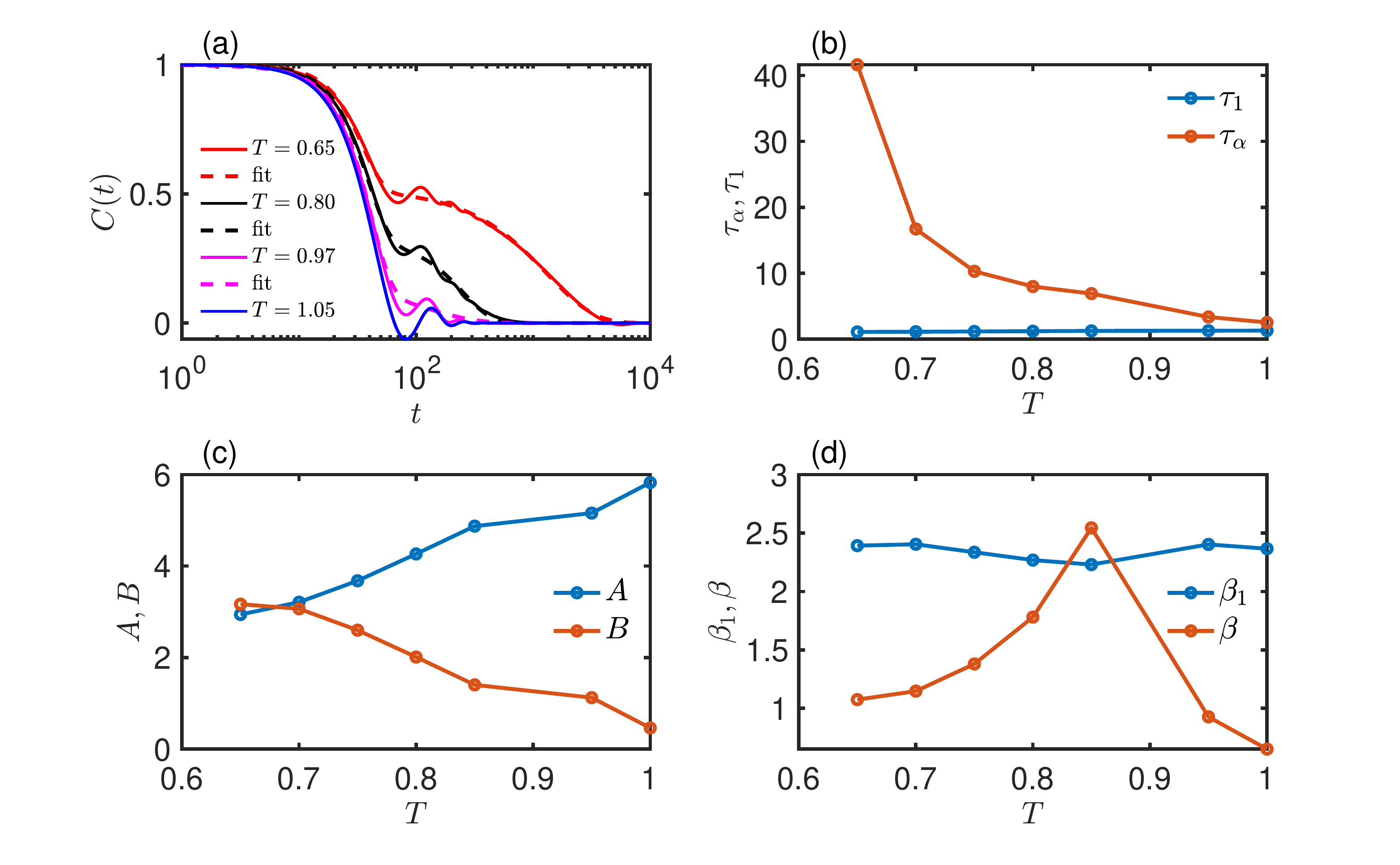}
	\caption{(a) The correlation function $C(t)$ (solid line) is shown as a function of time $(t)$ for several values of temperature in the classical limit, $\hbar=\sqrt{0.001}$. The fits to the correlation functions (dashed line with the same color) with the two-step relaxation function [Eq.\eqref{eq:TwoStep}] is also shown for $T=0.65, 0.80, 0.97 $, where $T=0.65$ is the closest to $T_d$.  (b) The behaviour of the relaxation times $\tau_1$ and $\tau_{\alpha}$ are shown as a function of  temperatures $T$. Though $\tau_1$ is almost constant, the $\alpha$-relaxation time ($\tau_{\alpha}$) diverges $T_d\simeq 0.62$ is approached. (c) The dependence of coefficients $A$ and $B$ are shown as a function of $T$. (d) The dependence of stretched exponents $\beta_1$ and $\beta$ are shown as a function of $T$.}
	\label{sup-fig-11}
\end{figure}

In this section we discuss the two-step glassy relaxation in the correlation function $C(t)$ in the PM phase above the dynamical transition temperature ($T_d$). In equilibrium, the correlation function can be obtained from retarded function using fluctuation-dissipation theorem. Using the fluctuation-dissipation theorem,
\beq 
C(\w) = \coth\left(\frac{\beta\w}{2}\right)\text{Im}Q^R(\w),
\eeq 
in the frequency domain. We obtain the correlation function $C(t)$ by doing Fourier transform, i.e. $C(t)=\int_{-\infty}^{\infty} (d\w/2\pi) e^{-\ci\w t} C(\w)$. A two-step relaxation or decay of correlation function is usually seen in classical glasses in the so-called $\beta$-relaxation regime above the glass transition temperature. The correlation functions initially has a faster decay to a plateau-like regime \cite{GotzeBook,Kob1997,Reichman2005}, where correlation decays with slow power laws, before eventually decaying as a stretched exponential in the $\alpha$-relaxation regime. In the quantum $p$-spin glass model, $C(t)$ also exhibits two-step relaxation as shown in Fig.\ref{fig:CorrelationFn}(c) (main text) and in Fig.\ref{sup-fig-11}(a). However, here the $\beta$-plateaus are mixed with oscillations, presumably due to quantum fluctuations and the existence of a soft gap (see Sec.\ref{supp-spect}) in the spectrum. To describe the overall decay profile of $C(t)$ [Fig.\ref{sup-fig-11}(a)], we take the following general two-step relaxation form 
\beq 
f(t) = A\exp\left[-(t/\tau_1)^{\beta_1}\right] + B\exp\left[-(t/\tau_{\alpha})^{\beta}\right]. \label{eq:TwoStep}
\eeq 
We fit numerically the correlation function $C(t)$ with the above function. The second stretched exponential describes the final $\alpha$-decay of the correlation function $C(t)$. We show the extracted fitting parameters $A,~B,~\tau_1,~\tau_\alpha,~\beta_1,~\beta$ in Figs.\ref{sup-fig-11}(b),(c) and (d) for a few temperatures for the classical limit $\hbar=\sqrt{0.001}$. We find that initial decay time $\tau_1$ more or less remains constant approaching $T_d$, whereas the $\alpha$-relaxation time $\tau_\alpha$ tends to diverge for $T\to T_d$ [Fig.\ref{sup-fig-11}(b)]. The contributions $A$ and $B$ for the two decays change with temperature with the $\alpha$-relaxation starting to dominate close to $T_d$ [Fig.\ref{sup-fig-11}(c)]. We find both the initial and later relaxations to be stretched exponentials as shown in Fig.\ref{sup-fig-11}(d). (Note that we denote the stretched exponent $\beta$ for the $\alpha$-relaxation following the standard notation \cite{Reichman2005} in the literature. This symbol should not be confused with inverse temperature `$\beta$'). 

As the correlation function has oscillations around the $\beta$ plateaus, so it is slightly tricky to obtain the onset temperature ($T_{\beta}$) of the two-step relaxation. We approximately estimate the $T_{\beta}$ from the temperature where the value of $C(t)$ at the first minimum of oscillation in the $\beta$ plateau turns negative to positive approaching from high temperature. In Fig.\ref{sup-fig-11}(a), we see that the first minimum of $C(t)$ for temperature $T\sim 0.97$ turns positive and then for all temperatures $T<0.97$, $C(t)$ remains positive. 
We thus estimate $T_{\beta}\approx 0.97$ for $\hbar=\sqrt{0.001}$. Similarly, we find $T_{\beta}$'s for other values of $\hbar$ and obtain $T_{\beta}(\hbar)$ curve as shown in Fig.\ref{fig:PhaseDiagram} in the main text. The two-step relaxation cannot be clearly distinguished for $\hbar\gtrsim 1.4$.

\section{Chaos from relaxation in PM phase}
In this section, we discuss about the nontrivial effect of complex relaxation in glasses on chaos near dynamical transition temperature ($T_d$). We numerically show that only the stretched exponential part $\sim \exp[-(t/\tau_{\alpha})^{\beta_a}]$ from the $\alpha$-relaxation of the correlation function $C(t)$ (shown in Fig. \ref{fig:CorrelationFn}a, main text) with the rapidly increasing $\tau_{\alpha}$ shown in Fig.\ref{fig:CorrelationFn}(b) (main text) can give rise to the non-monotonic $\lambda_L(T)$ and the broad maximum in it above $T_d$. For concreteness, we mainly focus on the classical limit $\hbar=\sqrt{0.001}$, where we find stretching exponent $\beta_a\simeq 1-2.5$ as shown in Fig.\ref{sup-fig-11}(c). We further show numerically that, instead of stretched exponential, even Debye exponential relaxation ($\beta_a=1$) with $\tau_{\alpha}$ shown in Fig.\ref{fig:CorrelationFn}(b) alone also gives rise to the non-monotonic $\lambda_L(T)$ and the broad maximum above $T_d$. Furthermore, we analytically derive the  non-monotonic behaviour of $\lambda_L(T)$ above $T_d$  using a Debye exponential decay in classical or high temperature limit.  
\subsection*{Numerical results}
\begin{figure}[H]
    \centering
    \includegraphics[width=0.5\textwidth]{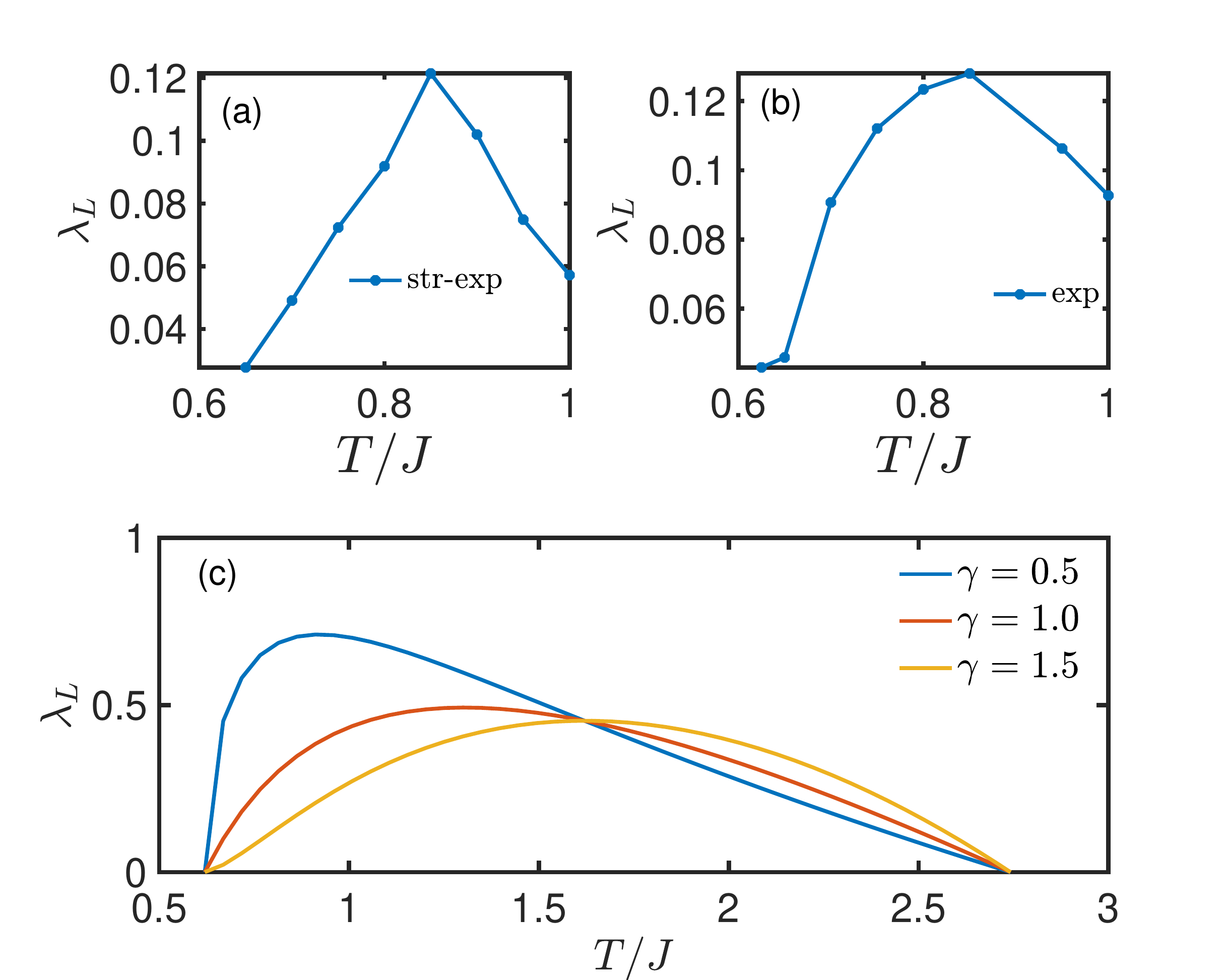}
    \caption{(a) Numerically computed Lyapunov exponent $\lambda_L(T)$ obtained from the stretched exponential part of the $\alpha$-relaxation. (b) The Debye exponential relaxation also gives rise to non-monotonic  $\lambda_L$ vs $T/J$. (c) The analytic result for non-monotonic behaviour of $\lambda_L(T)$ in Eq.\ref{eq:lambdaL_Debye} is plotted for $\gamma=0.5, 1.0, 1.5$ with $A_{\alpha}=0.55$. }
    \label{fig_S11}
\end{figure}

We first approximate the correlation function  as stretched exponential i.e
\beq
C(t) = A \exp[(-(|t|/\tau_{\alpha})^{\beta_a}]\label{eq:strExp}
\eeq 
where $A$ is a normalisation factor.
Then, we compute $C(\w)$ numerically in the frequency domain using fast Fourier transform (fft). The spectral function is obtained using FDT from $C(\w)$, i.e.
\begin{align} \label{FDT}
\text{Im}Q^R(\w) &= \tanh{(\frac{\w}{2T})}C(\w)  \\
\rho(\w) &= -\frac{1}{\pi}\text{Im}Q^R(\w) 
\end{align}
The normalisation factor $A$ is determined using the spherical constraint relation i.e $-\int^{\infty}_{\infty} d\w \rho(\w) n_B(T, \w) = 1$. We further calculate $Q_R(\pm \w+\ci \frac{\lambda_L}{2})$ via spectral representation of $Q_R(\w)$. Finally, we numerically solve the eigenvalue equation 
\ref{sup:fa} with kernel \ref{sup:Ka-w} for PM phase. The result for $\lambda_L(T)$, clearly exhibiting the broad maximum above $T_d$, is shown in Fig.\ref{fig_S11} (a). 

To verify that result is not specific to the stretched exponential form, we approximate the correlation function by Debye exponential relaxation, which is ubiquitous in liquids and many other systems, i.e 
\beq
C(t) = Ae^{-\Gamma_{\alpha}|t|},  \hspace{0.5cm} C(\w) = A\frac{\Gamma_{\alpha}}{\Gamma^2_{\alpha}+\w^2} \label{eq:Debye}
\eeq 
where $\Gamma_{\alpha}=1/\tau_{\alpha}$. Again, using FDT relation \ref{FDT}, we obtain the retarded function as   
\begin{align}
\text{Im}Q^R(\w) &= A\tanh{\bigg(\frac{\w}{2T}\bigg)}\frac{\Gamma_{\alpha}}{\Gamma^2_{\alpha}+\w^2} \label{eq:DebyeQ}\\
\rho(\w) &= -\frac{1}{\pi}\text{Im}Q^R(\w) 
\end{align}
The normalization constant ($A$) is fixed by imposing spherical constraint as earlier.  We use this spectral function to obtain $\lambda_L$ numerically from the eigenvalue equation \ref{sup:fa}. The result $\lambda_L(T)$ is shown in Fig.\ref{fig_S11} (b). This qualitatively gives the same result as that for the stretched exponential relaxation, namely a non-monotonic temperature dependence for $\lambda_\mathrm{L}$.

\subsection*{Analytical result}
In this section we derive an analytical result for $\lambda_\mathrm{L}$ for Debye exponential relaxation $\sim \exp(-t/\tau_\alpha)$ with a relaxation time $\tau_\alpha\sim (T-T_d)^\gamma$ ($\gamma>0$). Thus we establish an explicit relation between relaxation and chaos in the complex glassy relaxation regime where the relaxation time rapidly increases approaching the glass transition.. 

We solve the eigenvalue equation \eqref{sup:fa} for the kernel \eqref{sup:Ka-w} in the classical limit, where the retarded function $Q^R(\w)$ is obtained from $C(\w)$ [Eq.\eqref{eq:Debye}] using the classical FDT relation i.e. the high-temperature limit ($T\gg \w$) of Eq.\eqref{eq:DebyeQ},
\begin{align}
\text{Im}Q^R(\w) &= A\frac{\w}{2T}\frac{\Gamma_{\alpha}}{\Gamma^2_{\alpha}+\w^2},
\end{align}
and 
\begin{align}
Q^R(\w) &= A\frac{\Gamma_{\alpha}}{2T}\frac{1}{\Gamma_{\alpha}-\ci \w},
\end{align}
where $\Gamma_\alpha=1/\tau_\alpha=A_\alpha(T-T_d)^\gamma$, which goes to zero as we approach the dynamical transition temperature $T_d$. From numerical fit to $\tau_{\alpha}$ shown in Fig.\ref{fig:CorrelationFn}(b) (main text) with the above power law form, we obtain $\gamma=0.95\simeq 1$ and $A_{\alpha}=0.55$. Using $n_B(\w)=1/(e^{\beta\w}-1)\simeq T/\w$ for $T\gg \w$ in the spherical constraint, $A=2$.  
Hence, we have 
\begin{align}
Q^R(\pm \w + \ci \frac{\lambda_L}{2}) &= \frac{\Gamma_{\alpha}}{T}\frac{1}{\Gamma_{\alpha}+\frac{\lambda_L}{2}\mp \ci \w}
\end{align}
and 
\begin{align}
Q^W(\w) &= -\frac{\pi \rho(\w)}{\sinh(\w/2T)}\simeq 2\frac{\Gamma_{\alpha}}{\Gamma^2_{\alpha}+\w^2} \hspace{0.5cm} (T\gg \w).
\end{align}
From the eigenvalue Eq. \ref{sup:fa}, with kernel \ref{sup:Ka-w},
we get
\begin{align}
6J^2 \bigg(\frac{\Gamma_{\alpha}}{T}\bigg)^2 \frac{1}{(\Gamma_{\alpha}+\lambda_L/2)^2+\w^2} \int^{\infty}_{-\infty}  \frac{d\w'}{2\pi} \frac{\Gamma_{\alpha}}{\Gamma^2_{\alpha}+(\w-\w')^2}f(\w') = f(\w) 
\end{align}
Now to analytically solve the above equation for $\lambda_L$ and $f(\w)$, we make the following approximation of replacing a Lorentzian by a Gaussian, which is expected to give qualitatively the same result, 
\beq 
\frac{1}{\Gamma^2_{\alpha}+\w^2} \to \frac{1}{\Gamma^2_{\alpha}}\sqrt{\frac{\pi}{2}}e^{-\w^2/2\Gamma^2_{\alpha}}
\eeq 
The above approximation with appropriate normalization maintains the spherical constraint. Thus, we get  
\begin{align}
\frac{3J^2}{\pi}\bigg(\frac{\Gamma_{\alpha}}{T}\bigg)^2 \frac{\Gamma_{\alpha}}{\tilde{\Gamma}^2_{\alpha}\Gamma^2_{\alpha}}\frac{\pi}{2}e^{-\w^2/2\tilde{\Gamma}^2_{\alpha}}\int^{\infty}_{-\infty}d\w' e^{-(\w-\w')^2/2\Gamma^2_{\alpha}}f(\w')=f(\w)     
\end{align}
where $\tilde{\Gamma}_{\alpha}=\Gamma_{\alpha}+\lambda_L/2$. The above integral equation can be solved by using
\beq 
f(\w) = e^{-\w^2/2\sigma^2},
\eeq 
so that,
\begin{align}\label{eq:gauss}
\frac{3J^2}{2T^2}\frac{\Gamma_{\alpha}}{\tilde{\Gamma}^2_{\alpha}}e^{-\w^2/2\tilde{\Gamma}^2_{\alpha}}\int^{\infty}_{-\infty}d\w' e^{-(\w'-\w)^2/2\Gamma^2_{\alpha}}e^{-\w'^2/2\sigma^2}=e^{-\w^2/2\sigma^2}.
\end{align}
We clearly can see that the Gaussian form of the $f(\w)$ satisfies the above equation since the integral over $\w'$ above produces another Gaussian, provided
\begin{align}
\frac{1}{\Bar{\Gamma}^2_{\alpha}}&=\frac{1}{\tilde{\Gamma}^2_{\alpha}}+\frac{2}{\Gamma^2_{\alpha}}-\frac{\Bar{\Gamma}^2_{\alpha}}{\Gamma^4_{\alpha}} \\
\Bar{\Gamma}_{\alpha} & = \frac{\tilde{\Gamma}^2_{\alpha}}{a\Gamma_{\alpha}}
\end{align}
where $1/\bar{\Gamma}_{\alpha}^2=1/\Gamma_\alpha^2+1/\sigma^2$ and $a=c^2J^2/T^2$ with $c=(3\sqrt{2\pi}/2)^{1/2}\approx 2$. From the above, we obtain
\beq
\frac{1}{\tilde{\Gamma}^2_{\alpha}}+ \frac{2}{\Gamma^2_{\alpha}}-\frac{\tilde{\Gamma}^4_{\alpha}}{a^2\Gamma^6_{\alpha}} = \frac{a^2\Gamma^2_{\alpha}}{\tilde{\Gamma}^4_{\alpha}}
\eeq 
Solving the above equation for $\tilde{\Gamma}_\alpha$ for real roots, we finally obtain
\begin{align}
 \lambda_L &= 2\Gamma_{\alpha}\bigg[\frac{1}{\sqrt{2}}\sqrt{\bigg(\frac{cJ}{T}\bigg)^2 + \bigg(\frac{cJ}{T}\bigg)\sqrt{\bigg(\frac{cJ}{T}\bigg)^2+4} }-1\bigg]  \label{eq:lambdaL_Debye}
\end{align}
The above result for $\lambda_L$ is plotted in Fig.\ref{fig_S11} for different values of $\gamma$. The non-monotonic variation of $\lambda_L$ with $T$ and the broad maximum at a temperature $T_m>T_d$ is nicely captured by the analytical result [Eq.\eqref{eq:lambdaL_Debye}]. However, in this approximation, $\lambda_L\propto \Gamma_\alpha \to 0$ as $T\to T_d$. This is, of course, an artifact of approximating the two-step relaxation in Fig. \ref{fig:CorrelationFn}a (main text) by a single exponential with a diverging time scale. The same is true for the numerical results of the preceding subsection where we use a single stretched exponential or exponential decay for $C(t)$ [Eq.\eqref{eq:strExp}]. Nevertheless, the approximation captures the non-monotonic part of $\lambda_L(T)$.

We now analytically estimate $T_m$ assuming $cJ/T\gg 1$, where 
\beq 
\lambda_L \approx 2A_{\alpha}(T-T_d)^{\gamma}\bigg[ \frac{cJ}{T}-1\bigg]. 
\eeq 
From the above expression, it is clear that $\lambda_L$ initially increases for $T\gtrsim T_d$ due to the increase of $\Gamma_\alpha=1/\tau_\alpha$, but eventually $\lambda_L$ is expected to decrease due to crossover from strong ($J>T$) to weak ($J<T$) coupling at high temperature. In between, the maximum in $\lambda_L(T)$ appears due to competition between the increasing $1/\tau_\alpha$ and factor $(cJ/T-1)$ decreasing with temperature. 
The position ($T_m$) of maximum of $\lambda_L$ is found from $\partial \lambda_L/\partial T=0$, 
\beq
 T_m = \frac{cJ(\gamma-1)+\sqrt{c^2J^2(\gamma-1)^2+4\gamma cJ T_d}}{2\gamma}
\eeq 
For $\gamma \approx 1$, as estimated from the numerical fit to $\tau_{\alpha}$ shown in Fig.\ref{fig:CorrelationFn}(b) (main text), we get
 \beq
 T_m = \sqrt{cJ T_d}
\eeq 
which satisfies $T_d<T_m<cJ$ since $cJ>T_d$. Thus $T_m$ signifies a crossover in chaos, arising due to an interplay of relaxation, the rapid increase of relaxation time in the glassy regime, and the crossover from strong coupling ($J\gtrsim T$) to weak coupling ($J \lesssim T$). This result is more general than the model considered here and should have implications for complex relaxations in liquids and many other interacting systems, since exponential Debye relaxation is ubiquitous. As shown in Fig.\ref{fig:PhaseDiagram} (main text), we also observe a correlation between $T_m$ and the onset temperature $T_\beta$ for two-step relaxation. It is quite likely that the onset of non-trivial temperature dependence of $\tau_\alpha\sim (T-T_d)^{-\gamma}$, which is connected with the non-monotonic $\lambda_L(T)$, is correlated with the onset of the two-step relaxation too. However, to properly establish this relation, we need an independent understanding of $T_\beta$, which is beyond the scope of the current work and will be an interesting future direction of study.

\end{document}

%% file: definitions.tex
\newcommand{\usenomenclature}{}
\ifdefined\usenomenclature
\usepackage[noprefix]{nomencl}
\fi

\newcommand{\beq}{\begin{equation}}
\newcommand{\eeq}{\end{equation}}

\newcommand{\bcen}{\begin{center}}
\newcommand{\ecen}{\end{center}}
\newcommand{\btab}{\begin{tabular}}
\newcommand{\etab}{\end{tabular}}
\newcommand{\bdes}{\begin{description}}
\newcommand{\edes}{\end{description}}

\newcommand{\bea}{\begin{eqnarray}}
\newcommand{\eea}{\end{eqnarray}}

\newcommand{\bary}{\begin{array}}
\newcommand{\eary}{\end{array}}
\newcommand{\benum}{\begin{enumerate}}
\newcommand{\eenum}{\end{enumerate}}
\newcommand{\bitem}{\begin{itemize}}
\newcommand{\eitem}{\end{itemize}}

\makeatletter

\newcommand{\Rmnum}[1]{\expandafter\@slowromancap\romannumeral #1@}
\makeatother

\newcommand{\ci}{\mathfrak{i}}

\newcommand{\Tr}{\textup{Tr}}

\newcommand{\tbd}[1]{}

\newcommand{\concept}[1]{}
\newcommand{\figconcept}[1]{}
\newcommand{\eqnconcept}[1]{}
\newcommand{\tabconcept}[1]{}
\newcommand{\appconcept}[1]{}

\newcommand{\G}[2]{G^{#2}_{#1}}

\usepackage{cancel}

%% file: CrossoverQuantumToClassical.bbl
\begin{thebibliography}{53}
\bibitem [{foo()}]{footnote1}%
  \BibitemOpen
  \href@noop {} {}\bibinfo {note} {The classical limit $\Gamma\to0$ can be also
  taken by $M\to\infty$ with fixed $\hbar$, but this limit trivially
  corresponds to infinitely massive particle and there is no dynamics in the
  classical limit. For all our results, we calculate $\lambda_\mathrm{L}$ (in
  units of $J/\hbar$) as a function of $\Gamma$ and obtain $\lambda_\mathrm{L}$
  in units of $J$ by dividing with $\sqrt{\Gamma J}$ for $M=1$. The $\hbar\to
  0$ limit, unlike the $M\to\infty$ limit, leads to non-trivial classical
  dynamics, e.g. two-step glassy relaxation in the PM phase.}\BibitemShut
  {Stop}%
\end{thebibliography}
